\documentclass[12pt]{article}

\usepackage{CJKutf8}          
\usepackage{graphicx}         
\usepackage{epstopdf}
\usepackage{amsmath,amssymb}  
\usepackage{booktabs}         
\usepackage{longtable}        
\usepackage{authblk}
\usepackage{array}

\usepackage[a4paper,margin=1in]{geometry}

\newcommand{\keywords}[1]{\par\small\textbf{Keywords:} #1\par}

\title{Heterogeneous immune recovery after viral response through a dynamical model of feedback-driven persistence and clearance}
\author[1]{Xiaoxin Wang}
\author[1]{Kai Kang}
\author[1]{Leyi Zhang}
\author[1,*]{Changjing Zhuge}
\affil[1]{School of Mathematics Statistics and Mechanics, Beijing University of Technology, Beijing 100124, China}
\affil[*]{Correspondence: zhuge@bjut.edu.cn}
\date{}
\begin{document}
\begin{CJK*}{UTF8}{gbsn}

\maketitle

\begin{abstract}
Viral infections trigger complex immune responses with heterogeneous outcomes shaped by nonlinear feedbacks. An ordinary differential equation model is developed to investigate immune response dynamics during viral infection, incorporating six modules: viral load, innate immunity, cellular immunity, humoral immunity, immune suppression, and IL-6 levels. Bifurcation analysis reveals that under continuous viral exposure, when viral clearance rate and intrinsic viral death rate satisfy specific conditions, the system exhibits up to five stable equilibria. This indicates that different health and disease states may coexist depending on initial conditions, while severe inflammation mainly arises from strong activation of cellular immunity, highlighting the complexity of immune responses. Simulations of finite-time viral exposure demonstrate multi-timescale recovery characteristics: viral load and IL-6 levels decline rapidly, whereas humoral immune activation and immunosuppression show delayed and sustained patterns. Furthermore, analysis of infectious period and disease duration also indicates that during transition from early acute response to chronic disease, viral replication rate plays a critical role, while immune response intensity is sensitive to both viral clearance and immune self-activation. Subsystem analysis identifies the three-component subsystem of viral load, innate immunity, and cellular immunity as core drivers of bistability and oscillations, while humoral immunity, immune suppression, and IL-6 primarily modulate response amplitude and timing. This work establishes a theoretical framework for analyzing immune response and chronic risks through feedback dynamical modelling, providing insights for intervention strategies.

\end{abstract}

\keywords{Virus–immune dynamics, Mathematical modeling, Bifurcation, Multistability, Subsystem analysis}

\section{Introduction}
The host immune response triggered by viral infection is a highly complex and dynamic process, involving nonlinear and coupled interactions among multiple cytokines and regulatory mechanisms. These immune interactions not only determine the course and severity of infection but also profoundly influence immune homeostasis and tissue repair during recovery. Classical viral dynamics models have successfully captured the basic relationship between viral load and cell infection. However, given the high complexity of feedback regulation within the immune system, an integrative perspective is required to reflect the overall regulatory effects of immunity on viral dynamics \cite{perelson2002modelling}.
In recent years, increasing evidence has shown that multistability, sustained oscillations, and sensitivity to initial conditions are key dynamical features underlying the heterogeneity of post-infection immune states \cite{ferrell2021systems}. Excessive immune activation or dysregulated suppression may lead to chronic inflammation, disruption of immune rhythms, and even immunopathological damage. The bidirectional regulation between the pro-inflammatory cytokine IL-6 and immunosuppressive pathways plays a central role in limiting excessive inflammation and maintaining homeostasis, but under certain conditions, it may also drive the system into a pathological high-inflammation state. Nevertheless, many existing models remain focused on single or limited immune modules, with relatively little attention paid to the immune network as a whole—particularly regarding how humoral and cellular immunity interact, and how the pro-inflammatory cytokine IL-6 regulates immunosuppressive pathways to maintain immune homeostasis.

This study proposes a virus–immune interaction network model that integrates viral dynamics with key host immune processes, virus ($[V]$), innate immunity ($[I]$), cellular immunity ($[C]$), humoral immunity ($[H]$), immune suppression ($[S]$), and the pro-inflammatory cytokine IL-6\cite{pawelek2012modeling}. Structurally, the model employs Hill functions to provide a unified description of cross-module activation, inhibition, and saturation effects \cite{alon2019introduction}, capturing the concentration dependence of viral replication on the one hand and the nonlinear inhibition of viral load by multiple immune clearance pathways on the other. In addition, an explicit exogenous viral input function $\alpha(t)$ is introduced to simulate different infection scenarios, including persistent and transient exposures. This design preserves the biological interpretability of both variables and parameters.
Existing studies of within-host viral dynamics are generally based on the target-cell limited framework, which has established a standard modeling paradigm for describing viral infection, replication, and clearance of susceptible cells, and has enabled the estimation of key parameters through data fitting \cite{perelson1999mathematical}. In the context of acute infections, innate immunity is often incorporated to explain the decline and clearance phases of viral load \cite{baccam2006kinetics}. Further extensions include the incorporation of adaptive immunity, where cellular immunity (CTLs/NK cells) and humoral immunity (antibodies) jointly represent directed clearance and neutralization barriers. Representative models include the Hancioglu–Swigon–Clermont framework, the quantitative analysis by Pawelek et al. on the parallel actions of innate and adaptive immunity, and the model of Saenz et al. that couples infection processes with pathological phenotypes \cite{hancioglu2007dynamical, pawelek2012modeling, saenz2010dynamics}. Several reviews have noted that despite these advances, most studies still tend to extend the standard model by selectively adding one or two immune pathways, while the coupling between pro-inflammatory cytokines and suppressive pathways, as well as scenarios with exogenous viral input, remain relatively underexplored \cite{beauchemin2011review, ciupe2017host}.

Based on the proposed virus–immune interaction network model, we systematically investigate the effects of external viral input patterns, key process parameters, and immune module combinations on the overall dynamical behavior. We first examine system responses under conditions of persistent viral input, showing that different combinations of immune clearance efficiency and viral replication capacity may lead to steady-state transitions, multistability, or the emergence of new pathological states. Critical parameter thresholds and their associated steady-state changes are quantified through numerical simulations and bifurcation analysis. We then analyze the recovery dynamics under non-persistent viral input scenarios, and find that different immune modules exhibit distinct timescales during the post-clearance recovery process (\ref{fig:result1}), with lagged declines observed across modules. In addition, to quantitatively describe the temporal features of the infection and inflammation phases, we introduce two time-based indicators: “infectious duration” and “illness duration, ” and assess their sensitivity to immune regulatory mechanisms under multiparameter perturbations. Finally, through subsystem enumeration analysis, we systematically screened variable combinations while fixing partial modules at steady state, identifying the triplet of virus ($[V]$), innate immunity ($[I]$), and cellular immunity ($[C]$) as the core structure driving complex dynamical modes such as bistability and oscillations. In contrast, humoral immunity and the suppression–inflammation loop primarily regulate response amplitude and recovery synchrony.

In summary, this work reveals the mechanisms underlying multistability and oscillations in the immune network, while providing quantifiable temporal indicators for evaluating the efficiency of immune regulation. It not only advances the theoretical understanding of virus–immune system coupling, but also offers a scalable analytical framework and methodological basis for the design and parameter optimization of personalized immune intervention strategies.

\section{Methods}

Significant differences exist in immune responses among individuals following viral infection, and these differences are closely associated with clinical manifestations. To elucidate the pathogenic mechanisms of viral infection \cite{wilk2020single, hadjadj2020impaired} and to provide theoretical insights for therapeutic strategies, we developed a virus–immune interaction model based on ordinary differential equations, with a particular focus on the overall regulatory role of the immune system. The model adopts a modular structure to represent the interactions among virus, innate immunity, adaptive immunity, immune suppression, and inflammatory cytokines, aiming to highlight the nonlinear couplings across multiple feedback loops. Steady-state and bifurcation analyses are employed to reveal the possible dynamical behaviors of the system. In contrast to recent approaches using spatially explicit or hybrid dynamical models to explore immune spatial features \cite{cai2023spatial}, this study emphasizes module coupling and stability analysis at the global level.

The human immune system is broadly divided into two components, innate immunity and adaptive immunity \cite{murphy2016janeway, sompayrac2022immune}. Upon viral invasion, innate immunity serves as the first line of defense, rapidly eliminating pathogens through physical and chemical barriers (e.g., skin, gastric acid), phagocytic cells (e.g., macrophages and neutrophils), and natural killer (NK) cells. This response is accompanied by an increase in pro-inflammatory factors such as IL-6, which establishes the inflammatory and signaling background for subsequent responses. Adaptive immunity is then activated to achieve targeted clearance, providing long-lasting protection against specific antigens \cite{alberts2002molecular}. Within adaptive immunity, T cells play a central role: CD8\textsuperscript{+} T cells recognize and destroy host cells infected by viruses or other pathogens, while CD4\textsuperscript{+} T cells coordinate interactions among immune cells and assist in their activation. Humoral immunity relies on B cells and their differentiation into plasma cells, which produce antibodies to neutralize pathogens and prevent further damage.
At the same time, to avoid excessive immune activation, suppressive pathways provide negative feedback on inflammation. Regulatory T cells (Tregs) inhibit overactive T cells, while immune checkpoint proteins (e.g., PD-1, CTLA-4) and anti-inflammatory cytokines (e.g., IL-10, $TGF-\beta$) also play critical roles in dampening interactions among immune cells. In the present model, these mechanisms are incorporated as interactions among the corresponding immune modules.

In the regulation of the immune system, humoral immunity also plays an important role. Humoral immunity influences the host’s defense capacity against pathogens by modulating immune cell functions and the strength of interactions among immune cells. Hormones such as cortisol, sex hormones (estrogen and testosterone), and thyroid hormones can regulate immune cell interactions by either promoting or suppressing immune cell activity and differentiation. Cortisol exerts a significant immunosuppressive effect by attenuating the strength of immune cell interactions, whereas estrogen generally enhances such interactions, particularly by promoting antibody production through B cells \cite{cain2017immune, trigunaite2015suppressive}. The regulatory role of humoral immunity in immune cell interactions is important not only for adaptive immunity but also for the functioning of the innate immune system.
The bidirectional regulation of IL-6 and suppressive mechanisms plays a key role in limiting excessive inflammation, maintaining homeostasis, and coordinating temporal responses. IL-6 is a major pro-inflammatory cytokine \cite{tang2020cytokine} that promotes immune cell activation and differentiation, especially in antibody production and T-cell function. It also plays a central role in the acute-phase response by stimulating hepatic synthesis of C-reactive protein (CRP) and enhancing local inflammatory reactions. Elevated IL-6 levels are closely associated with the development and progression of various diseases, including infections, cancers, and autoimmune disorders. Notably, IL-6 not only drives inflammatory responses but can also induce the activation of immunosuppressive pathways, thereby establishing a dynamic balance between pro-inflammatory and suppressive effects. This dual role makes it a critical node in regulating infection outcomes and immune homeostasis.

\begin{figure}[tbhp]
\begin{center}
\includegraphics[width=\textwidth]{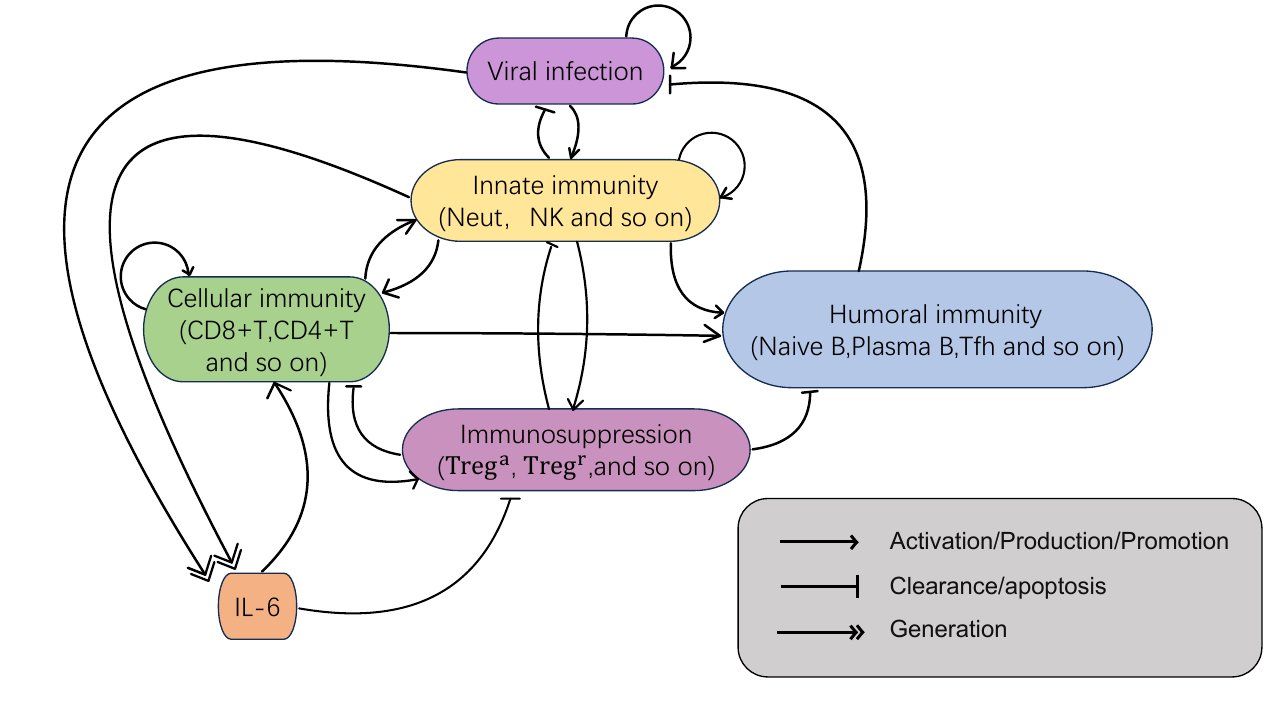}
\end{center}

\caption{Immune response network to viral infection. The network consists of six modules: virus, innate immunity, cellular immunity, humoral immunity, immune suppression, and interleukin-6(IL-6). Each module involves complex interactions between immune cells and cytokines. The innate immunity module mainly includes neutrophils and NK cells; the cellular immunity module is primarily composed of CD8\textsuperscript{+} T cells and CD4\textsuperscript{+} T cells; humoral immunity mainly involves naïve B cells, plasma B cells, and Tfh cells; and the immune suppression module is mainly composed of Treg cells. In this model, single arrows represent activation, production, or promotion of downstream components; bar-ended lines denote clearance or apoptosis processes; and double arrows indicate generation or differentiation into subsequent components.}\label{fig:1}
\end{figure}

To effectively characterize the host immune response mechanisms following viral invasion, we construct an interaction network of viral infection and immune response comprising six variables that represent different immune modules (Fig.~\ref{fig:1}). This network accounts for the complex interactions among viral dynamics ($[V]$), innate immunity ($[I]$), cellular immunity ($[C]$), humoral immunity ($[H]$), immune suppression ($[S]$), and interleukin-6 (IL-6) after infection, thereby capturing the global dynamical features of the host immune response following viral entry.

In the model, the dynamic evolution of viral load results from the combined effects of multiple mechanisms. First, the virus may enter the host system through external sources. To capture this effect, we introduce a time-dependent input function $\alpha(t)$. During a finite initial period $[0, T]$, the function is maintained at a constant value $\phi$, representing continuous exogenous viral exposure; thereafter, it drops to zero, indicating that external infection sources cease to contribute. In this way, $\alpha(t)$ characterizes the environmental exposure faced by the host at different stages. The mathematical definition of $\alpha(t)$ is provided in Eq. \eqref{eqV1}.

\begin{equation}
\alpha(t) =
\begin{cases}
\phi, & t\in[0, T], \\[0.5ex]
0, & t\in[T, \infty).\label{eqV1}
\end{cases}
\end{equation}

In addition to exogenous input, the virus itself is capable of replication. The proliferation rate of the virus does not increase linearly but instead exhibits a nonlinear saturating behavior as the concentration rises. To capture this concentration dependence, the model employs a Hill function (as shown in Eq. \eqref{eqV2}) to characterize the kinetics of viral growth.
\begin{equation}
    a_{v0}\frac{[V]^{n_{v0}}}{k_{v0}^{n_{v0}}+[V]^{n_{v0}}}\label{eqV2}
\end{equation}

This expression not only reflects the trend that the viral replication rate increases with viral abundance, but also reveals its saturating nature constrained by resources, environmental factors, or host conditions \cite{alon2019introduction}. Here, $a_{v0}$ denotes the maximum replication rate of the virus, i.e., the limiting speed under saturation; $n_{v0}$ is the Hill coefficient, which determines the steepness of the process and reflects the sensitivity of replication rate to changes in viral concentration; and $k_{v0}$ is the half-saturation constant, corresponding to the viral concentration at which the replication rate reaches half of its maximum value. In other words, this term represents not only a kinetic approximation but also a biological abstraction of the viral self-replication process.
At the same time, the host immune system continually participates in viral clearance, as innate immunity, cellular immunity, and humoral immunity all accelerate viral elimination. Specifically, virus death induced by cellular immunity is described by Eq. \eqref{eqV3}.
\begin{equation}
    \Bigg(d_{v1}\frac{[C]^{n_{v1}}}{k_{v1}^{n_{v1}}+[C]^{n_{v1}}}\Bigg)[V]\label{eqV3}
\end{equation}
Similarly, viral clearance mediated by innate immunity and humoral immunity can be represented by Eq. \eqref{eqV4}.
\begin{equation}
    \Bigg(d_{v2}\frac{[I]^{n_{v2}}}{k_{v2}^{n_{v2}}+[I]^{n_{v2}}}
+ d_{v3}\frac{[H]^{n_{v3}}}{k_{v3}^{n_{v3}}+[H]^{n_{v3}}}\Bigg)[V]\label{eqV4}
\end{equation}

In this expression, each term corresponds to the clearance effect of a specific immune mechanism: $dv_1$, $dv_2$, and $dv_3$ respectively quantify the clearance efficiencies of cellular immunity, innate immunity, and humoral immunity against the virus. The associated $n$ and $k$ parameters characterize the sensitivity and half-saturation level of these immune effects. The multiplication by $[V]$ reflects that the viral clearance rate depends not only on the strength of immune responses but also on the viral load itself.
Finally, even in the absence of immune action or exogenous input, viruses gradually decline due to intrinsic inactivation, environmental instability, and passive removal. This natural decay effect is modeled by $d_{v4}[V]$, serving as a background dissipative mechanism.
By integrating these four components—exogenous input, self-replication, immune clearance, and natural decay—we obtain the complete expression for viral dynamics as shown in Eq. \eqref{eqV5}.
\begin{equation}
\begin{aligned}
\dfrac{\mathrm{d}[V]}{\mathrm{d}t} 
&= \alpha(t) 
+ a_{v0}\frac{[V]^{n_{v0}}}{k_{v0}^{n_{v0}}+[V]^{n_{v0}}} \\
&\quad -\Bigg(
    d_{v1}\frac{[C]^{n_{v1}}}{k_{v1}^{n_{v1}}+[C]^{n_{v1}}}
  + d_{v2}\frac{[I]^{n_{v2}}}{k_{v2}^{n_{v2}}+[I]^{n_{v2}}} 
  + d_{v3}\frac{[H]^{n_{v3}}}{k_{v3}^{n_{v3}}+[H]^{n_{v3}}}
  \Bigg)[V]- d_{v4}[V]
\end{aligned}
\label{eqV5}
\end{equation}

Innate immunity ($[I]$) describes the rapid defensive response of the host during the early stage of viral invasion. Its dynamical evolution integrates four processes: viral activation, self-amplification of immunity, negative feedback regulation by suppressive mechanisms, and natural decay of immune factors.
First, the presence of the virus directly triggers the activation of innate immunity, which exhibits a saturation effect. This effect is represented using a concentration-dependent Hill function, as shown in Eq. \eqref{eqI1}.
\begin{equation}
    a_{I0}\frac{[V]^{n_{I0}}}{k_{I0}^{n_{I0}}+[V]^{n_{I0}}}\label{eqI1}
\end{equation}

Here, $a_{I0}$ denotes the maximum activation rate triggered by viral stimulation; $n_{I0}$ is the Hill coefficient, characterizing the steepness of the innate immune response to changes in viral level; and $k_{I0}$ is the half-saturation constant, corresponding to the viral load required for immune activation to reach half of its maximum level. This process reflects the sensitivity of host immune cells or molecules to viral stimulation during the early stage of infection.
Second, the innate immune system possesses a certain degree of self-amplification. For example, interferon molecules, once activated, can further enhance their own production, thereby forming a positive feedback loop. This effect can be represented by Eq. \eqref{eqI2}.
\begin{equation}
    a_{I1}\frac{[I]^{n_{I1}}}{k_{I1}^{n_{I1}}+[I]^{n_{I1}}}\label{eqI2}
\end{equation}

In Eq. \eqref{eqI2}, $a_{I1}$ represents the maximum self-activation rate, while $n_{I1}$ and $k_{I1}$ respectively determine the sensitivity of the response and the half-saturation point. This term reflects that, once initiated, innate immunity can be rapidly amplified through positive feedback, thereby strengthening its ability to suppress the virus. At the same time, the immune system must also avoid excessive reactions. To this end, the model introduces an immune suppression module ($[S]$), which suppresses innate immunity via negative feedback. This regulation is formulated as shown in Eq. \eqref{eqI3}.
\begin{equation}
    d_{I2}\frac{[S]^{n_{I2}}}{k_{I2}^{n_{I2}}+[S]^{n_{I2}}}[I]\label{eqI3}
\end{equation}

In this equation, $d_{I2}$ represents the strength of the suppressive effect, while $n_{I2}$ and $k_{I2}$ control the nonlinear threshold characteristics of the process. Through this mechanism, the model reflects that host defense, while being activated, is also subject to regulation in order to prevent excessive immune responses that could cause tissue damage. Finally, innate immune factors undergo natural death or degradation, represented by the term $d_{I3}[I]$, which indicates that immune molecules or cells gradually decay in the absence of continuous stimulation.
By combining Eqs. \eqref{eqI1}, \eqref{eqI2}, and  \eqref{eqI3}, the dynamics of innate immunity can be summarized in the form of Eq. \eqref{eqI4}. This equation comprehensively characterizes the activation, amplification, regulation, and decay mechanisms of innate immunity, reflecting the rapid and dynamic defensive features exhibited by the host during the early stage of infection.
\begin{equation}
\dfrac{\mathrm{d}[I]}{\mathrm{d}t} = a_{I0}\frac{[V]^{n_{I0}}}{k_{I0}^{n_{I0}}+[V]^{n_{I0}}}
+a_{I1}\frac{[I]^{n_{I1}}}{k_{I1}^{n_{I1}}+[I]^{n_{I1}}}
-d_{I2}\frac{[S]^{n_{I2}}}{k_{I2}^{n_{I2}}+[S]^{n_{I2}}}[I]
-d_{I3}[I]\label{eqI4}
\end{equation}  

During viral infection, cellular immunity ($[C]$) primarily represents the T cell–mediated specific immune response. Its core function is to recognize and eliminate host cells infected by the virus, thereby blocking viral replication and transmission \cite{wherry2022t}. The dynamics of this process can be decomposed into four aspects: initiation, amplification, regulation, and death. First, the activation of cellular immunity requires a “dual condition, ” namely the presence of a sufficient viral load ($[V]$) together with the prior activation of innate immunity ($[I]$). Accordingly, Eq. \eqref{eqC1} effectively characterizes the early activation process of cellular immunity.
\begin{equation}
    a_{C0}\frac{[I]^{n_{C0}}}{k_{C0}^{n_{C0}}+[I]^{n_{C0}}}\frac{[V]^{n_{C1}}}{k_{C1}^{n_{C1}}+[V]^{n_{C1}}}\label{eqC1}
\end{equation}

In this equation, $a_{C0}$ denotes the maximum initiation rate, while $n_{C0}$, $k_{C0}$, $n_{C1}$, and $k_{C1}$ respectively control the sensitivity and threshold of cellular immunity to the levels of innate immunity and viral concentration. The multiplicative form of this term clearly reflects the fine regulation of the immune system, ensuring that a specific cellular immune response is effectively initiated only when the viral concentration exceeds a certain level and innate immunity has been activated.
Second, once cellular immunity is initiated, a positive feedback mechanism is also present \cite{kaech2012transcriptional}. Activated T cells or effector cells can enhance their own expansion and maintenance through cytokine-mediated processes. This effect is represented in the form of a Hill function, as shown in Eq. \eqref{eqC2}.

\begin{equation}
    a_{C2}\frac{[C]^{n_{C2}}}{k_{C2}^{n_{C2}}+[C]^{n_{C2}}}\label{eqC2}
\end{equation}

$a_{C2}$ represents the maximum self-activation rate, while $n_{C2}$ and $k_{C2}$ respectively determine the sensitivity and half-saturation level of the response. This term reflects that, once initiated, cellular immunity can form a sustained amplification effect, maintaining a high level of viral clearance ability over an extended period. Meanwhile, cellular immunity is also regulated by immune suppressive factors ($[S]$). The negative feedback mechanism governing this regulation is represented by Eq. \eqref{eqC3}.
\begin{equation}
    d_{C3}\frac{[S]^{n_{C3}}}{k_{C3}^{n_{C3}}+[S]^{n_{C3}}}[C]\label{eqC3}
\end{equation}
 
In this equation, $d_{C3}$ represents the suppression intensity, and $n_{C3}$ and $k_{C3}$ characterize the nonlinear features of this process. This term reflects the limitation of cellular immunity activity by immune suppressive factors, preventing excessive immune responses that could cause immune-related tissue damage. Finally, effector cells of cellular immunity also undergo natural apoptosis or inactivation, a process represented by the term $d_{C4}[C]$, indicating that even in the absence of suppressive factors, effector cells such as T cells gradually decline.
Therefore, by integrating the above equations  \eqref{eqC1}, \eqref{eqC2}, and  \eqref{eqC3}, along with the natural apoptosis of cellular immunity, the dynamics of cellular immunity in viral infection can be represented by Eq. \eqref{eqC4}. This equation systematically encapsulates the entire process of cellular immunity during viral infection, driven by both innate immunity and the virus, including activation, self-amplification, negative feedback regulation by suppressive mechanisms, and eventual natural decay.
\begin{equation}
\dfrac{\mathrm{d}[C]}{\mathrm{d}t}=a_{C0}\frac{[I]^{n_{C0}}}{k_{C0}^{n_{C0}}+[I]^{n_{C0}}}\frac{[V]^{n_{C1}}}{k_{C1}^{n_{C1}}+[V]^{n_{C1}}}
+a_{C2}\frac{[C]^{n_{C2}}}{k_{C2}^{n_{C2}}+[C]^{n_{C2}}}
-d_{C3}\frac{[S]^{n_{C3}}}{k_{C3}^{n_{C3}}+[S]^{n_{C3}}}[C]
-d_{C4}[C]\label{eqC4}
\end{equation}

In the host immune response, humoral immunity ($[H]$) primarily represents the adaptive immune response mediated by B cells. Compared to innate immunity and cellular immunity, the activation of humoral immunity typically occurs later during infection; however, it can effectively clear free viral particles through the specific recognition and neutralization by antibodies \cite{rastogi2022role}. Its dynamics integrate activation signals from cellular and innate immunity, are regulated by immune suppression mechanisms, and are accompanied by the natural decay of humoral immunity itself.
One major source of activation for humoral immunity is cellular immunity ($[C]$). T cells not only directly kill infected cells but also promote B cell differentiation and antibody production through cytokine secretion. This process is represented by Eq. \eqref{eqH1}.
\begin{equation}
    a_{H0}\frac{[C]^{n_{H0}}}{k_{H0}^{n_{H0}}+[C]^{n_{H0}}}\label{eqH1}
\end{equation}

$a_{H0}$ represents the maximum facilitation rate of cellular immunity in activating humoral immunity, while $n_{H0}$ and $k_{H0}$ respectively determine the nonlinear steepness and half-saturation point of the response. This term reflects the synergistic relationship between cellular and humoral immunity, where T cells provide critical support for B cell activation and antibody production.
Furthermore, innate immunity ($[I]$) can also provide activation signals for humoral immunity. For example, interferon molecules or inflammatory factors can indirectly enhance B cell function. This effect is represented by Eq. \eqref{eqH2}.
\begin{equation}
    a_{H1}\frac{[I]^{n_{H1}}}{k_{H1}^{n_{H1}}+[I]^{n_{H1}}}\label{eqH2}
\end{equation}

In this equation, $a_{H1}$ represents the maximum facilitation rate of innate immunity in activating humoral immunity, while $n_{H1}$ and $k_{H1}$ control the dynamics of this process. This term reflects the cascade relationship between different immune modules, enabling humoral immunity to be effectively activated under the influence of multiple signals. Meanwhile, humoral immunity is also regulated by negative feedback from the immune suppression module ($[S]$), as shown in Eq. \eqref{eqH3}.
\begin{equation}
    d_{H2}\frac{[S]^{n_{H2}}}{k_{H2}^{n_{H2}}+[S]^{n_{H2}}}[H]\label{eqH3}
\end{equation}

In this equation, $d_{H2}$ represents the suppression intensity parameter, while $n_{H2}$ and $k_{H2}$ determine the threshold characteristics of this effect. This term reflects that, while activating humoral immunity, the immune system also prevents excessive antibody responses through suppressive mechanisms, thereby avoiding unnecessary damage to the host. Finally, humoral immune molecules (such as antibodies) undergo a natural degradation process, represented by the term $d_{H3}[H]$, which indicates that, even in the absence of viral stimulation, antibody levels gradually decline.
Therefore, the dynamics of humoral immunity can be summarized in the form of Eq. \eqref{eqH4}, which comprehensively reflects the key role of humoral immunity in adaptive immune responses. Specifically, it is initiated by the combined activation signals from cellular and innate immunity, exerts neutralizing effects of specific antibodies under the regulation of immune suppression, and gradually decays after the response ends.
\begin{equation}
\dfrac{\mathrm{d}[H]}{\mathrm{d}t} = a_{H0}\frac{[C]^{n_{H0}}}{k_{H0}^{n_{H0}}+[C]^{n_{H0}}}
+a_{H1}\frac{[I]^{n_{H1}}}{k_{H1}^{n_{H1}}+[I]^{n_{H1}}}
-d_{H2}\frac{[S]^{n_{H2}}}{k_{H2}^{n_{H2}}+[S]^{n_{H2}}}[H]
-d_{H3}[H]\label{eqH4} 
\end{equation}

In the immune system, it is not possible for all modules to maintain a high state continuously. Therefore, immune suppression regulates this by introducing a negative feedback mechanism through the immune suppression module ($[S]$), which reflects the host’s effort to prevent overactivation of immune responses. This module plays a crucial role in maintaining immune homeostasis and preventing tissue damage. The dynamics of this process are influenced by activation and regulation from cellular immunity ($[C]$), innate immunity ($[I]$), and inflammatory factor IL-6, while also being limited by its own natural decay.
First, immune suppression has a baseline level, represented by the constant $\delta$. In the model, $\delta=0.0001$, which signifies that even in the absence of external stimuli, the host maintains a certain degree of immune suppression to ensure homeostasis. Second, under external stimulation, both cellular immunity ($[C]$) and innate immunity ($[I]$) can promote the production of immune suppressive factors. This process is represented by Eq. \eqref{eqS1}.
\begin{equation}
    a_{S0}\frac{[C]^{n_{S0}}}{k_{S0}^{n_{S0}}+[C]^{n_{S0}}}+a_{S1}\frac{[I]^{n_{S1}}}{k_{S1}^{n_{S1}}+[I]^{n_{S1}}}\label{eqS1}
\end{equation}

In this equation, $a_{S0}$ and $a_{S1}$ represent the maximum rates at which cellular immunity and innate immunity activate immune suppressive factors, respectively, while $n_{S0}$, $k_{S0}$ and $n_{S1}$, $k_{S1}$ control the nonlinear characteristics of their responses. This part reflects that when the immune system is active, the body simultaneously enhances immune suppression to avoid excessive immune responses. In addition, the inflammatory factor IL-6, at high levels, further drives the suppressive effect, forming an inflammation-suppression negative feedback loop, as shown in Eq. \eqref{eqS2}.
\begin{equation}
    d_{S2}\frac{[IL6]^{n_{S2}}}{k_{S2}^{n_{S2}}+[IL6]^{n_{S2}}}[S]\label{eqS2}
\end{equation}

$d_{S2}$ represents the intensity parameter for this feedback effect, while $n_{S2}$ and $k_{S2}$ characterize its threshold and nonlinear effects. This term reflects that when inflammation increases, the body attempts to enhance the suppressive response in order to reduce inflammatory damage.
Finally, immune suppressive factors undergo natural decay or inactivation, represented by $d_{S3}[S]$. This term indicates that, even in the absence of continuous activation, the level of immune suppression gradually declines.

In summary, the dynamics of the immune suppression module can be expressed by Eq. \eqref{eqS3}. This equation incorporates both the baseline level and the upregulation effect from immune activation, while also considering the negative feedback regulation driven by inflammatory factors and the natural decay of suppressive factors, thereby maintaining a dynamic balance between activation and suppression in the immune system.
\begin{equation}
\dfrac{\mathrm{d}[S]}{\mathrm{d}t} = \delta
+ a_{S0}\frac{[C]^{n_{S0}}}{k_{S0}^{n_{S0}}+[C]^{n_{S0}}}
+ a_{S1}\frac{[I]^{n_{S1}}}{k_{S1}^{n_{S1}}+[I]^{n_{S1}}}
- d_{S2}\frac{[IL6]^{n_{S2}}}{k_{S2}^{n_{S2}}+[IL6]^{n_{S2}}}[S]
- d_{S3}[S]\label{eqS3}
\end{equation}

Interleukin-6 (IL-6) is an important pro-inflammatory cytokine produced during viral infection, widely regarded as a key marker of immune system activation and inflammation levels \cite{tang2020cytokine}. The dynamic changes in its levels not only reflect the intensity of the host’s immune response to infection but are also closely related to inflammation-associated pathological conditions. In the model, the dynamics of IL-6 are primarily regulated by viral load, cellular immunity levels, and its own degradation process.First, cellular immunity ($[C]$) is one of the key drivers of IL-6 production. Activated T cells and associated immune factors promote the secretion of IL-6, a process represented by Eq. \eqref{eqIL1}.
\begin{equation}
    a_{IL0}\frac{[C]^{n_{IL0}}}{k_{IL0}^{n_{IL0}}+[C]^{n_{IL0}}}\label{eqIL1}
\end{equation}

$a_{IL0}$ represents the maximum activation rate, while $n_{IL0}$ and $k_{IL0}$ describe the sensitivity and half-saturation level of this process, respectively. This term reflects the pro-inflammatory effect associated with active cellular immunity.
Additionally, viral load ($[V]$) itself can directly stimulate the production of IL-6, especially during infection spread and inflammation escalation. This process is described by Eq. \eqref{eqIL2}.
\begin{equation}
    a_{IL1}\frac{[V]^{n_{IL1}}}{k_{IL1}^{n_{IL1}}+[V]^{n_{IL1}}}\label{eqIL2}
\end{equation}
 
In this equation, $a_{IL1}$ represents the maximum induction rate of IL-6 production by the virus, while $n_{IL1}$ and $k_{IL1}$ determine the nonlinear characteristics of the response. This term reflects the positive correlation between viral load and inflammation levels.
Finally, IL-6 molecules undergo a natural degradation process, represented by $d_{IL2}[IL6]$. Even in the absence of stimulation, IL-6 gradually declines to maintain homeostasis. Therefore, the dynamics of IL-6 can be uniformly expressed by Eq. \eqref{eqIL3}.
\begin{equation}
\dfrac{\mathrm{d}[IL6]}{\mathrm{d}t} = a_{IL0}\frac{[C]^{n_{IL0}}}{k_{IL0}^{n_{IL0}}+[C]^{n_{IL0}}}
+a_{IL1}\frac{[V]^{n_{IL1}}}{k_{IL1}^{n_{IL1}}+[V]^{n_{IL1}}}
-d_{IL2}[IL6]\label{eqIL3}
\end{equation}

This equation comprehensively reflects the key role of IL-6 in infection and inflammatory responses. It is driven by both cellular immunity and viral load, while being limited by its own degradation. A sustained increase in IL-6 levels typically indicates that the host has entered a state of high inflammation, which may lead to severe pathological reactions and clinical symptoms.

By combining the interactions between the different immune modules described above (Eqs. \eqref{eqV4}, \eqref{eqI4}, \eqref{eqC4}, \eqref{eqH4}, \eqref{eqS3}, and  \eqref{eqIL3}), the dynamics of each immune module in the human body after viral infection are modeled using equations based on the widespread use of Hill functions \cite{alon2019introduction}. The following equation \ref{model_equation} is constructed.

\begin{eqnarray}
    \dfrac{\mathrm{d}[V]}{\mathrm{d}t}&=&\alpha(t)+a_{v0}\frac{[V]^{n_{v0}}}{k_{v0}^{n_{v0}}+[V]^{n_{v0}}}-\Bigg(d_{v1}\frac{[C]^{n_{v1}}}{k_{v1}^{n_{v1}}+[C]^{n_{v1}}}+d_{v2}\frac{[I]^{n_{v2}}}{k_{v2}^{n_{v2}}+[I]^{n_{v2}}}\nonumber\\
    &&+d_{v3}\frac{[H]^{n_{v3}}}{k_{v3}^{n_{v3}}+[H]^{n_{v3}}}\Bigg)[V]-d_{v4}[V]\\
    \dfrac{\mathrm{d}[I]}{\mathrm{d}t}&=&a_{I0}\frac{[V]^{n_{I0}}}{k_{I0}^{n_{I0}}+[V]^{n_{I0}}}+a_{I1}\frac{[I]^{n_{I1}}}{k_{I1}^{n_{I1}}+[I]^{n_{I1}}}-d_{I2}\frac{[S]^{n_{I2}}}{k_{I2}^{n_{I2}}+[S]^{n_{I2}}}[I]-d_{I3}[I]\\
    \dfrac{\mathrm{d}[C]}{\mathrm{d}t}&=&a_{C0}\frac{[I]^{n_{C0}}}{k_{C0}^{n_{C0}}+[I]^{n_{C0}}}\frac{[V]^{n_{C1}}}{k_{C1}^{n_{C1}}+[V]^{n_{C1}}}+a_{C2}\frac{[C]^{n_{C2}}}{k_{C2}^{n_{C2}}+[C]^{n_{C2}}}-
    d_{C3}\frac{[S]^{n_{C3}}}{k_{C3}^{n_{C3}}+[S]^{n_{C3}}}[C]\nonumber\\
    &&-d_{C4}[C]\\
    \dfrac{\mathrm{d}[H]}{\mathrm{d}t}&=&a_{H0}\frac{[C]^{n_{H0}}}{k_{H0}^{n_{H0}}+[C]^{n_{H0}}}+a_{H1}\frac{[I]^{n_{H1}}}{k_{H1}^{n_{H1}}+[I]^{n_{H1}}}-d_{H2}\frac{[S]^{n_{H2}}}{k_{H2}^{n_{H2}}+[S]^{n_{H2}}}[H]-d_{H3}[H]\\
    \dfrac{\mathrm{d}[S]}{\mathrm{d}t}&=&\delta+a_{S0}\frac{[C]^{n_{S0}}}{k_{S0}^{n_{S0}}+[C]^{n_{S0}}}+a_{S1}\frac{[I]^{n_{S1}}}{k_{S1}^{n_{S1}}+[I]^{n_{S1}}}-d_{S2}\frac{[IL6]^{n_{S2}}}{k_{S2}^{n_{S2}}+[IL6]^{n_{S2}}}[S]-d_{S3}[S]\\
    \dfrac{\mathrm{d}[IL6]}{\mathrm{d}t}&=&a_{IL0}\frac{[C]^{n_{IL0}}}{k_{IL0}^{n_{IL0}}+[C]^{n_{IL0}}}+a_{IL1}\frac{[V]^{n_{IL1}}}{k_{IL1}^{n_{IL1}}+[V]^{n_{IL1}}}-d_{IL2}[IL6]
\end{eqnarray}\label{model_equation}

\subsection{Parameters and numerical solution}

To better capture the heterogeneity of host immune responses during viral infection, both in terms of intensity and timing, and to account for the resulting diversity of clinical manifestations, we systematically explored the parameter space of the virus–immune response interaction network and performed corresponding dynamical simulations.
To balance exploration capacity with computational efficiency, parameter-space sampling was implemented. The model equations include six variables and 63 parameters, making it impractical to fully cover the high-dimensional parameter space; therefore, the sampling dimensions and sample size were reduced to improve computational efficiency. Specifically, the half-saturation constants and Hill coefficients, which characterize system dynamical features, were fixed and not sampled. In bifurcation analyses, random sampling was applied to major parameters, including the viral replication rate, activation and clearance rates among immune compartments, decay rates of immune compartments, and cytokine production rates. For each parameter set, initial values of the variables were assigned, and the system of ordinary differential equations was numerically solved using the \texttt{ode15s} solver in MATLAB. Furthermore, we also examined how different viral input patterns influence the dynamical trajectories of immune state variables.

In addition, to ensure that the model computations capture the qualitative characteristics of the system, appropriate initial conditions and parameter values were selected based on existing immunological studies. Specifically, the initial conditions specify the concentrations of virus, immune cells, and cytokines set at the beginning of the simulations. These concentrations represent the baseline state of the immune system, i.e., the normal levels in the absence of viral infection or other external perturbations. Parameter values, in contrast, characterize the key mechanisms of immune responses, including how the virus infects host cells and how immune cells are activated or suppressed. These values were chosen according to the known characteristics of immune responses. For example, they include the strength of interactions among immune cells following viral invasion, the action modes of suppressive cytokines, and how cytokines regulate immune responses \cite{carsetti2020different, janeway2001course}. 
The specific parameter settings are listed in Table~\ref{parameter}.

\begin{small}
\renewcommand{\arraystretch}{1.1}
\begin{longtable}{p{2cm} >{\raggedright\arraybackslash}p{7.5cm} p{2cm}}
\caption{Model parameters and descriptions} \label{parameter} \\
\toprule
Parameter & Description & Value \\ 
\midrule[1.5pt]
\endfirsthead

\multicolumn{3}{c}{{Continuation of Table \ref{parameter}}} \\
\toprule
Parameter & Description & Value \\
\midrule[1.5pt]
\endhead

\endfoot

\endlastfoot

$\alpha(t)$  & Exogenous viral input rate; constant $\phi$ for $t\in[0, T]$, then 0 & 0.01 \\ \hline
$a_{v0}$     & Maximum viral replication rate & 1.25 \\ \hline
$k_{v0}$     & Half-saturation constant of viral replication & 0.2 \\ \hline
$n_{v0}$     & Hill coefficient of viral replication & 3 \\ \hline
$d_{v1}$     & Maximum clearance strength of virus by cellular immunity & 0.12 \\ \hline
$k_{v1}$     & Half-saturation constant for clearance of virus by cellular immunity & 0.1 \\ \hline
$n_{v1}$     & Hill coefficient for clearance of virus by cellular immunity & 3 \\ \hline
$d_{v2}$     & Maximum clearance strength of virus by innate immunity  & 0.08 \\ \hline
$k_{v2}$     & Half-saturation constant for clearance of virus by innate immunity & 0.225 \\ \hline
$n_{v2}$     & Hill coefficient for clearance of virus by innate immunity & 3 \\ \hline
$d_{v3}$     & Maximum clearance strength of virus by humoral immunity & 0.03 \\ \hline
$k_{v3}$     & Half-saturation constant for clearance of virus by humoral immunity & 0.3 \\ \hline
$n_{v3}$     & Hill coefficient for clearance of virus by humoral immunity  & 3 \\ \hline
$d_{v4}$     & Natural decay rate of virus & 0.25 \\ \hline

$a_{I0}$     & Maximum activation rate of innate immunity induced by virus & 0.15 \\ \hline
$k_{I0}$     & Half-saturation constant for activation of innate immunity by virus & 0.025 \\ \hline
$n_{I0}$     & Hill coefficient for activation of innate immunity by virus & 3 \\ \hline
$a_{I1}$     & Maximum self-activation rate of innate immunity & 0.25 \\ \hline
$k_{I1}$     & Half-saturation constant for self-activation of innate immunity & 0.1 \\ \hline
$n_{I1}$     & Hill coefficient for self-activation of innate immunity & 3 \\ \hline
$d_{I2}$     & Suppression strength of innate immunity by immune suppression & 0.4 \\ \hline
$k_{I2}$     & Half-saturation constant for suppression of innate immunity by immune suppression & 3 \\ \hline
$n_{I2}$     & Hill coefficient for suppression of innate immunity by immune suppression & 3 \\ \hline
$d_{I3}$     & Natural decay rate of innate immunity & 0.8 \\ \hline

$a_{C0}$     & Maximum co-activation of cellular immunity induced jointly by innate immunity and virus & 0.8 \\ \hline
$k_{C0}$     & Half-saturation constant for innate immunity in co-activation of cellular immunity & 0.05 \\ \hline
$n_{C0}$     & Hill coefficient for innate immunity in co-activation of cellular immunity & 3 \\ \hline
$k_{C1}$     & Half-saturation constant for virus in co-activation of cellular immunity & 0.05 \\ \hline
$n_{C1}$     & Hill coefficient for virus in co-activation of cellular immunity & 3 \\ \hline
$a_{C2}$     & Maximum self-activation rate of cellular immunity & 0.5 \\ \hline
$k_{C2}$     & Half-saturation constant for self-activation of cellular immunity & 0.4 \\ \hline
$n_{C2}$     & Hill coefficient for self-activation of cellular immunity & 3 \\ \hline
$d_{C3}$     & Suppression strength of cellular immunity by immune suppression & 0.4 \\ \hline
$k_{C3}$     & Half-saturation constant for suppression of cellular immunity by immune suppression & 3 \\ \hline
$n_{C3}$     & Hill coefficient for suppression of cellular immunity by immune suppression & 3 \\ \hline
$d_{C4}$     & Natural decay rate of cellular immunity & 0.4 \\ \hline

$a_{H0}$     & Maximum activation rate of humoral immunity  induced by cellular immunity & 0.15 \\ \hline
$k_{H0}$     & Half-saturation constant for activation of humoral immunity  by cellular immunity & 0.05 \\ \hline
$n_{H0}$     & Hill coefficient for activation of humoral immunity  by cellular immunity & 3 \\ \hline
$a_{H1}$     & Maximum activation rate of humoral immunity  induced by innate immunity & 0.05 \\ \hline
$k_{H1}$     & Half-saturation constant for activation of humoral immunity  by innate immunity & 0.05 \\ \hline
$n_{H1}$     & Hill coefficient for activation of humoral immunity  by innate immunity & 3 \\ \hline
$d_{H2}$     & Suppression strength of humoral immunity  by immune suppression & 1.25 \\ \hline
$k_{H2}$     & Half-saturation constant for suppression of humoral immunity  by immune suppression & 0.4 \\ \hline
$n_{H2}$     & Hill coefficient for suppression of humoral immunity  by immune suppression & 3 \\ \hline
$d_{H3}$     & Natural decay rate of humoral immunity  & 0.25 \\ \hline

$\delta$     & Basal generation rate of immune suppression & 0.0001 \\ \hline
$a_{S0}$     & Activation rate of immune suppression induced by cellular immunity & 0.05 \\ \hline
$k_{S0}$     & Half-saturation constant for activation of immune suppression by cellular immunity & 0.08 \\ \hline
$n_{S0}$     & Hill coefficient for activation of immune suppression by cellular immunity & 3 \\ \hline
$a_{S1}$     & Activation rate of immune suppression induced by innate immunity & 0.05 \\ \hline
$k_{S1}$     & Half-saturation constant for activation of immune suppression by innate immunity & 0.08 \\ \hline
$n_{S1}$     & Hill coefficient for activation of immune suppression strength by innate immunity & 3 \\ \hline
$d_{S2}$     & Down-regulation strength of immune suppression strength by IL-6 (negative feedback) & 0.4 \\ \hline
$k_{S2}$     & Half-saturation constant for down-regulation of immune suppression strength by IL-6 & 0.1 \\ \hline
$n_{S2}$     & Hill coefficient for down-regulation of immune suppression strength by IL-6 & 3 \\ \hline
$d_{S3}$     & Natural decay rate of immune suppression strength & 0.4 \\ \hline

$a_{IL0}$    & Maximum production rate of IL-6 induced by cellular immunity & 0.0125 \\ \hline
$k_{IL0}$    & Half-saturation constant for IL-6 production induced by cellular immunity & 0.5 \\ \hline
$n_{IL0}$    & Hill coefficient for IL-6 production induced by cellular immunity & 3 \\ \hline
$a_{IL1}$    & Maximum production rate of IL-6 induced by virus & 0.5 \\ \hline
$k_{IL1}$    & Half-saturation constant for IL-6 production induced by virus & 0.025 \\ \hline
$n_{IL1}$    & Hill coefficient for IL-6 production induced by virus & 3 \\ \hline
$d_{IL2}$    & Natural decay rate of IL-6 & 1.5 \\ 

\end{longtable}
\end{small}

\section{Results}
\subsection{Characteristic dynamical behavior}

\begin{figure}[tbhp]
\begin{center}
\includegraphics[width=\textwidth]{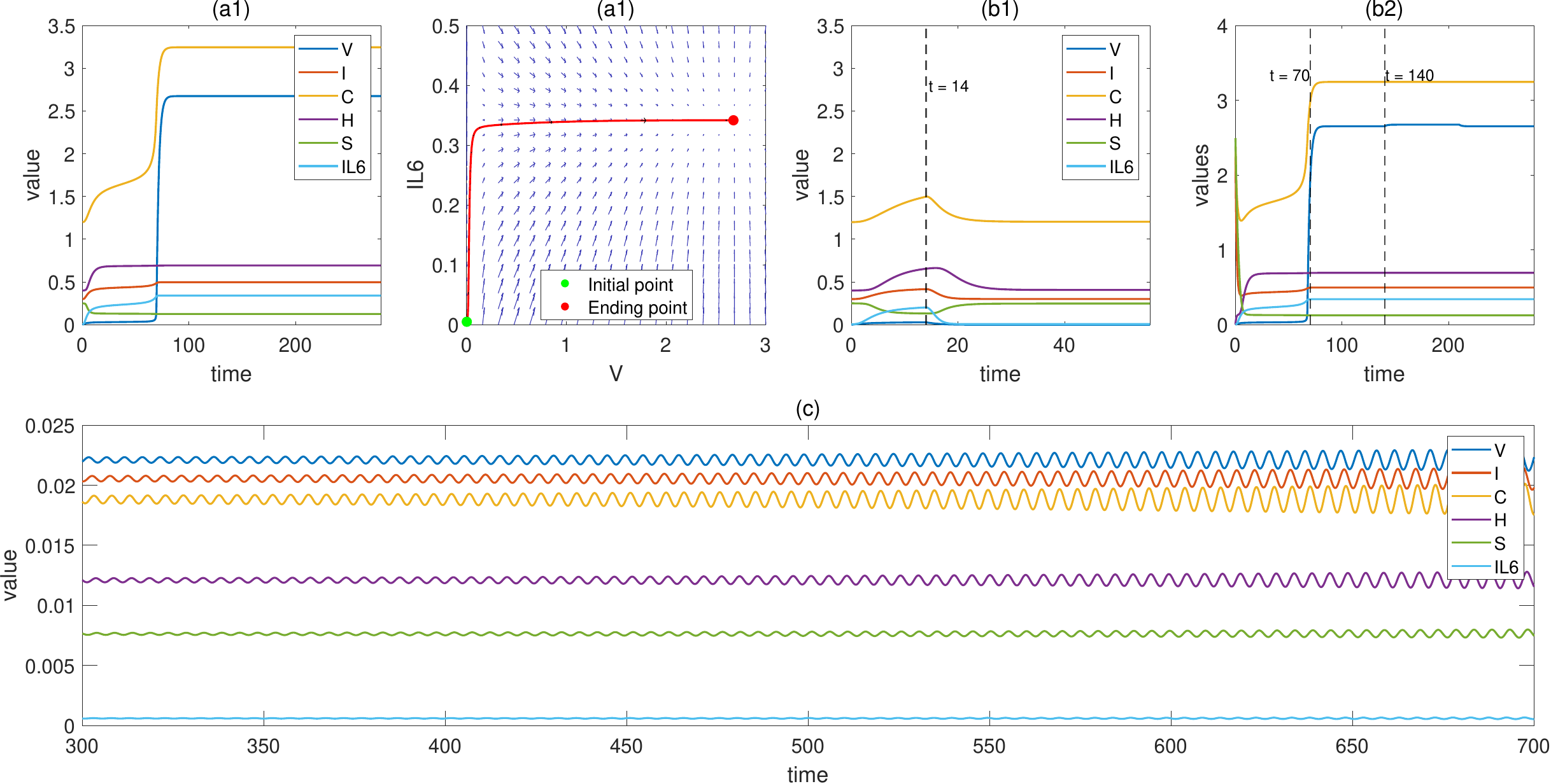}
\end{center}
\caption{\textbf{(a1)} Under continuous external viral input, the system variables ($[V](t)$, $[I](t)$, $[C](t)$, $[H](t)$, $[S](t)$, $IL-6(t)$) evolve over time into the typical response trajectories shown in the figure, eventually approaching a stable state at the final state.
\textbf{(a2)} The convergence process from the initial state (green point) to the steady state (red point). It illustrates the vector field of the system variables involved in (a1). The direction of the arrows indicates the local evolutionary trend in the phase space.
\textbf{(b1)} Immune response under short-term viral input ($t \in [0, 14]$), where some variables gradually stabilize after viral clearance.
\textbf{(b2)} After short-term viral input ($t \in [0, 70]$) stops, input is resumed at $t \in [140, 210]$ and then stops again. Compared to result \textbf{(b1)}, after viral clearance, the variables settle into a new steady state.
\textbf{(c)} Immune oscillatory state generated under a specific set of initial conditions and parameter combinations, where the system variables exhibit stable periodic oscillations, demonstrating that the model can exhibit complex steady-state oscillatory behavior under certain configurations.}\label{fig:result1}
\end{figure}

Based on the above setup, to elucidate the dynamical interplay between viral replication and immune responses, we performed numerical simulations to examine system evolution under different conditions and selected representative outcomes of virus–immune interactions. Fig.~\ref{fig:result1}(a1)–(a2) illustrates the dynamics of immune modules under continuous external viral input. During the early phase of sustained exposure, immune modules are rapidly activated, while the suppression module decreases, collectively resisting viral invasion. Although the immune modules can transiently restrain viral growth at first, viral expansion arises not only from external input but also from self-replication, which is inherently nonlinear. The half-saturation property of the Hill function limits unbounded viral proliferation, yet under persistent external input, even a small influx may push the system beyond a latent threshold. Once this threshold is crossed, the existing immune modules become insufficient to control viral growth, leading to a rapid rise in viral load. In response, immune modules adjust accordingly, and the system eventually converges to a new steady state. This phenomenon suggests that the host exhibits a latent period under sustained viral exposure, after which viral load escalates rapidly. These results indicate that external viral input can trigger an irreversible transition of immune system states, reflecting the system’s multistability.

Under short-term external input, the system returns to its original steady state after the input is removed (see Fig.~\ref{fig:result1}(b1)), indicating that a brief viral exposure can induce transient fluctuations in the immune system, but the overall structure remains stable without state transition. In contrast, when the input duration is sufficiently long (e.g., applied for $70, \mathrm{units}$ before removal), the system trajectory is pushed into another basin of attraction and eventually converges to a new steady state. Subsequently, when viral input is applied again during $t \in [140, 210]$, the input no longer alters the system’s steady state (see Fig.~\ref{fig:result1}). This demonstrates that although viral input is removed after a period of time, the accumulated effect has already driven the system to a new steady state, such that later removal or reintroduction of input produces no significant change in the system’s stability.

A more special case is shown in Fig.~\ref{fig:result1}(c), which illustrates that under certain initial conditions and parameter configurations, the system may enter a sustained oscillatory state. In this case, variables such as virus load $[V]$,innate immunity $[I]$, cellular immunity$[C]$, and humoral immunity $[H]$ all exhibit stable periodic oscillations. This dynamical pattern typically indicates that the immune system, after viral input, is driven into a critical regime, where the interplay of positive and negative feedback among different immune factors leads the system into periodic fluctuations. It reflects both the sensitivity of the system to viral input and its inherent nonlinear dynamics, and may correspond to pathological phenomena such as chronic inflammation or immune rhythm disorder.

\subsection{Bifurcation behavior of the system}

\begin{figure}[tbhp]
\begin{center}
\includegraphics[width=0.9\textwidth]{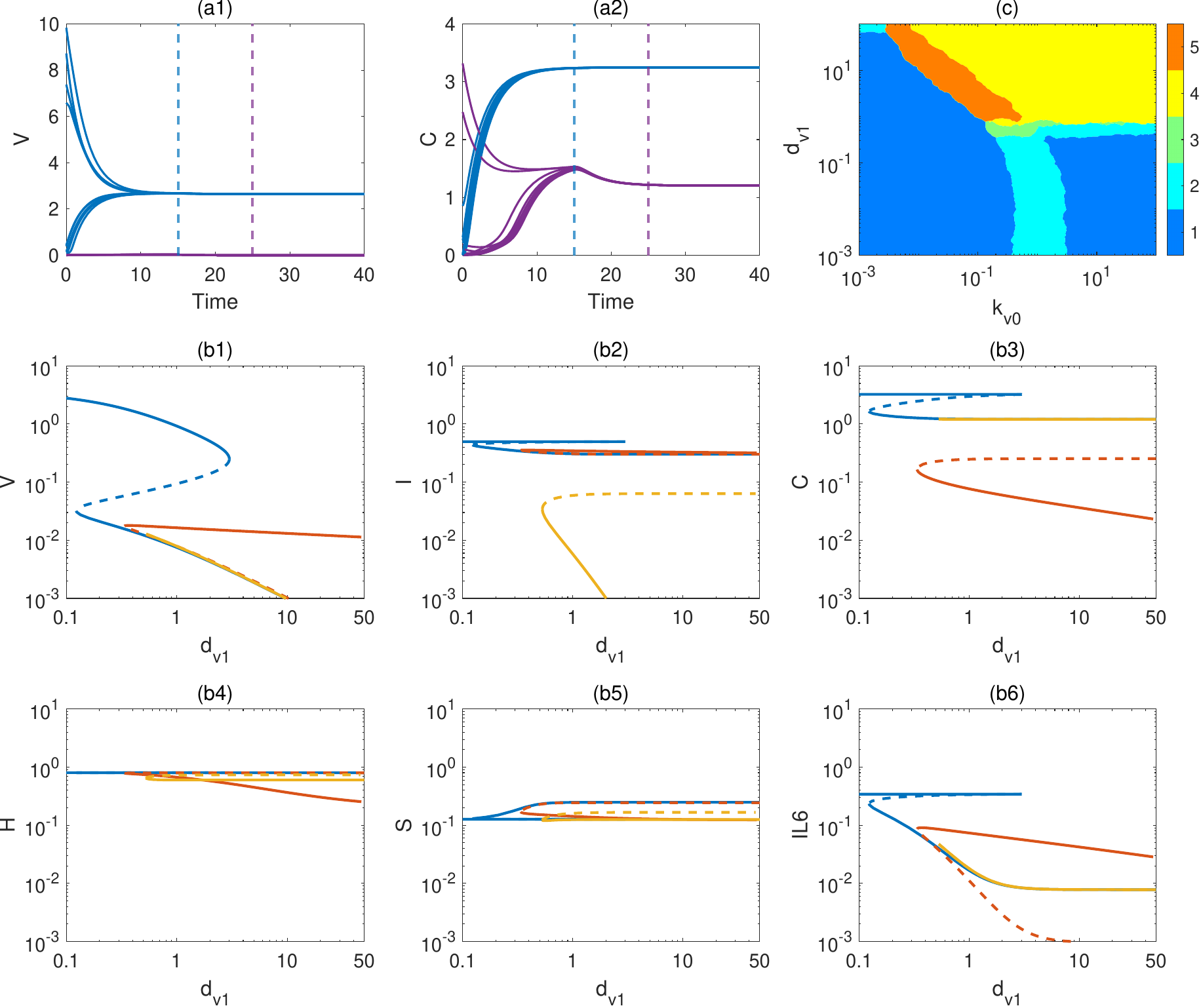}
\end{center}
\caption{The steady-state transitions of the system under continuous viral input are shown, illustrating how equilibrium states respond to variations in $d_{v1}$ (the clearance efficiency of cellular immunity against the virus).\textbf{(a1)} Time trajectories of viral load $[V](t)$ under multiple initial conditions; \textbf{(a2)} Corresponding trajectories of cellular immunity $[C](t)$; the blue and purple dashed lines mark the times at which the system reaches steady states, $t_1=15$ and $t_2=25$, respectively. A negative correlation between the response time and response strength of cellular immunity can be observed.  
\textbf{(b1)–(b6)}  The bifurcation structures of the six state variables. Solid lines denote stable branches, dashed lines denote unstable branches, and different colors correspond to distinct solution branches. Multiple saddle-node bifurcations occur, leading to the creation and disappearance of branches and forming regions of multistability. As $d_{v1}$ increases, the viral load $[V](t)$ and pro-inflammatory cytokine IL-6 decrease significantly, while other immune modules ($[I](t), [C](t), [H](t), [S](t)$) display coordinated regulation. This indicates that enhanced cellular immune clearance reshapes both the number and distribution of possible steady states, thereby determining alternative immune outcomes, although multistability persists.  
\textbf{(c)}  The parameter diagram. The number of steady states across combinations of $k_{v0}$ (viral production rate) and $d_{v1}$ (clearance rate). Both axes are shown in logarithmic scale. The color indicates the number of steady states (blue represents monostability, orange represents 5 steady states), revealing typical multistable regions and their boundaries.}\label{fig:result2}
\end{figure}

To analyze the dynamical characteristics of the system under continuous external viral input, we first consider the case of $\alpha(t)=\phi$. The results show that, with parameters fixed, the system can converge to different steady states solely by varying the initial conditions. As shown in Figs.~\ref{fig:result2}(a1)–(a2), panel (a1) depicts the time trajectories of viral load $V(t)$ under multiple initial conditions, while panel (a2) shows the trajectories of cellular immunity $C(t)$. The black and red dashed lines indicate the times when the system reaches steady states, $t_1=15$ and $t_2=25$, respectively, demonstrating a negative correlation between the response time and the response intensity of cellular immunity \cite{chatterjee2022modeling}. It can also be seen that although variable innate immunity $[I]$ ultimately converges to the same steady state, cellular immunity $[C]$ reaches two distinct stable states. Furthermore, under continuous viral input ($\phi = 0.01$), to examine the influence of key parameters on the system’s final state, we performed a bifurcation analysis of the model while keeping the initial conditions fixed. Figs.~\ref{fig:result2}(b1)–(b6) display the branch structures of stable and unstable solutions as key parameters vary. Different colors represent different steady-state solution branches, with solid lines denoting stable solutions and dashed lines denoting unstable solutions, clearly revealing the existence of multistability regions and their dynamic pathways.

It is particularly noteworthy that the parameter $d_{v1}$ in the model equation specifically represents the efficiency of cellular immunity in clearing the virus, reflecting the key impact of changes in the effectiveness of the cellular immune mechanism on the viral infection system. From the results shown in the figures, it is clearly observed that as the cellular immunity clearance ability ($d_{v1}$) changes, the system exhibits significant multistability and bifurcation behavior.
Fig.~\ref{fig:result2}(b1)–(b6) illustrates the multistability and bifurcation features of the system regulated by $d_{v1}$, underscoring the pivotal role of cellular immunity in shaping global immune steady states. Specifically, when the clearance ability of cellular immunity is weak (small $d_{v1}$), the viral steady-state level remains high, accompanied by elevated levels of the inflammatory cytokine IL-6, corresponding to clinical scenarios such as chronic viral infection or impaired host immune function. As $d_{v1}$ gradually increases, the enhanced clearance efficiency of cellular immunity drives a marked reduction in viral steady-state levels. Crossing one or more critical bifurcation points, the system transitions from a high-viral steady state to a low-viral steady state, entering a non-infectious healthy state. This process reflects a recovery pattern in which enhanced immune capacity or therapeutic intervention progressively clears the virus.
Furthermore, near the bifurcation points, the system is highly sensitive to initial conditions: different initial immune levels can lead to drastically different disease progression trajectories, manifested as either a long-term chronic infection state or a rapid recovery state with viral clearance. Specifically, under the same parameter conditions, both high viral-high inflammation steady states and low viral-low inflammation steady states coexist. At this point, the host’s initial immune level (e.g., initial immune cell count, baseline inflammatory response) will determine which steady state the system ultimately converges to. This phenomenon suggests that even with the same treatment, individuals may experience entirely different disease outcomes due to variations in their initial immune conditions.

To further elucidate the joint regulatory effects of viral replication and clearance mechanisms on the global stability structure of the system, this study extends the single-parameter bifurcation analysis by selecting two key dynamical factors: the viral replication rate parameter $a_{v0}$ and the viral clearance rate parameter $d_{v1}$. A two-dimensional parameter space scan was conducted to systematically evaluate their influence on the multistability structure of the model \cite{eftimie2016mathematical} (see Fig.~\ref{fig:result2}(c)).

Specifically, all other parameters were fixed, and a wide-range scan of $a_{v0}$ and $d_{v1}$ was performed in double-logarithmic coordinates. For each parameter combination, the number of steady states and their stability properties were determined by solving the steady-state system and performing eigenvalue analysis. The colors in the figure represent the number of steady states, ranging from monostability (blue) to five steady states (orange-red), showing a distinct regional distribution as illustrated in Fig.~\ref{fig:result2}(c).
It can be observed that in the region where $a_{v0}$ is small and $d_{v1}$ is large, the system possesses only one stable solution, corresponding to a healthy state in which the virus is effectively controlled by the immune system. When the viral replication rate increases (larger $a_{v0}$) and the clearance efficiency decreases (smaller $d_{v1}$), multiple steady states emerge, indicating that the system dynamics may fall into different infection outcomes, including viral persistence and uncontrolled immune activation.
In the transitional regions, rich tri-stability and multistability structures are observed, suggesting strong sensitivity to initial conditions and nonlinear response characteristics. In particular, in the orange regions, the system exhibits five steady states, reflecting highly complex steady-state topologies that may be accompanied by abundant bifurcation phenomena and unpredictable dynamical responses.
This two-parameter scan not only reveals the global steady-state distribution of the system under different combinations of viral replication and clearance intensities but also further emphasizes the nonlinear complexity of virus–host immune interactions. The results suggest that, within certain parameter ranges, the system may exhibit multiple potential courses of infection, where even small differences in the host’s initial immune state or intervention strategies can lead to completely different clinical outcomes. Therefore, this finding highlights the importance of individualized prediction and treatment of infectious diseases.

\subsection{coexisting high and low steady states}

\begin{figure}[tbhp]
\begin{center}
\includegraphics[width=\textwidth]{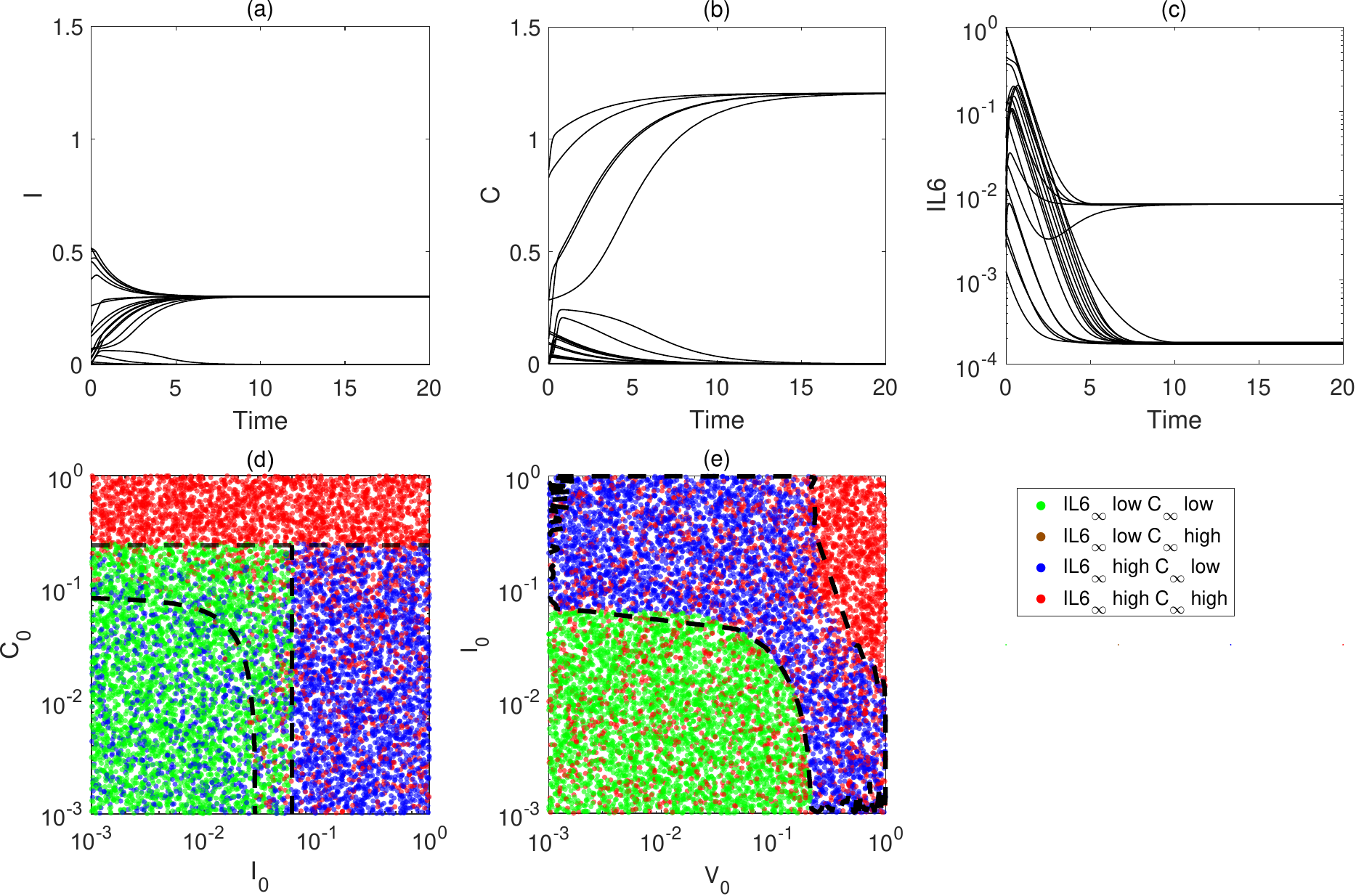}
\end{center}
\caption{Emergence of high and low states. \textbf{(a)–(c)} The temporal evolution of system states under 20 different initial conditions, all of which eventually converge to steady states. \textbf{(d)–(e)} The results of initial condition scans corresponding to different steady states. In panel \textbf{(d)}, the horizontal axis represents the initial values of innate immunity $[I](t)$, and the vertical axis represents the initial values of cellular immunity $[C](t)$. In panel  \textbf{(e)}, the horizontal axis represents the initial values of viral load $[V](t)$, and the vertical axis represents the initial values of innate immunity $[I](t)$. Different colors in the figures indicate the final steady states to which the system converges: green denotes low viral load $[V](t)$–high cellular immunity$[C](t)$, blue denotes high viral load $[V](t)$–low cellular immunity$[C](t)$, and red denotes other cases. The black dashed line indicates the division among different combinations. These results demonstrate that the system may converge to different steady states under varying initial conditions, highlighting the presence of multistability and sensitivity to initial conditions.}\label{fig:result4}
\end{figure}

In Fig.~\ref{fig:result2}(c), the steady states change very rapidly for certain parameter values. When $d_{v1}$ is large (e.g., $d_{v1}=100$), multiple changes in steady states occur as $k_{v0}$ increases. In fact, immune responses during viral infection vary significantly among individuals and are influenced by factors such as age, health status, and sex \cite{moore2020cytokine, guan2020comorbidity}. To further investigate the stochasticity and robustness of the system, several representative parameter points in specific regions were selected for analysis.
As shown in Fig.~\ref{fig:result4}(a)–(c), when $\alpha(t)$ is continuously applied and $d_{v1}$ is large, while keeping all other parameters unchanged, the temporal dynamics of viral load $[V]$, innate immunity $[I]$, cellular immunity $[C]$, and interleukin-6 (IL-6) were examined. The results indicate that, due to the relatively large value of the viral natural degradation coefficient $d_{v4}$, the viral load rapidly decreases to very low levels, or even close to zero, across all initial conditions. This outcome is consistent with the biological expectation of rapid viral clearance. However, other immune modules ($[I]$, $[C]$, IL-6) exhibit different trajectories. In particular, within the cellular immunity module, one trajectory remains at a high state, while another stays near zero, corresponding to a low state. We refer to these two scenarios as the high state and low state of cellular immunity \cite{kadelka2019modeling}.

When cellular immunity is in the high state, immune responses (such as T-cell activity) serve as a direct mechanism against viral infection. In this case, cellular immunity is highly active, allowing the immune system to effectively recognize and eliminate virus-infected host cells through T cells or other effector cells. Although the viral load is $V=0$, cellular immunity $[C]$ and the inflammatory cytokine IL-6 remain in the high state. This may reflect a “lag effect” of the immune system \cite{xie2010analysis} or the self-sustaining nature of immune responses. During viral clearance, the high states of cellular immunity $[C]$ and IL-6 may not be entirely virus-driven but instead maintained through their own positive feedback loops. Sustained high expression of cellular immunity $[C]$ can further activate and maintain innate immunity $[I]$ at relatively high levels, forming a positive feedback cycle of immune activation. In addition, as a key inflammatory mediator, IL-6 is directly or indirectly activated by cellular immunity $[C]$, and through positive feedback further enhances immune cell activity. Such positive feedback mechanisms may trap the system in a prolonged high-expression state. Thus, even when the viral load is zero, an excessively activated immune system can still induce inflammation. These high-expression states typically occur shortly after infection and represent part of the acute-phase immune response. Without external intervention, patients may remain in chronic pathological conditions, such as persistent inflammation or immune overactivation.
In contrast, when cellular immunity is in the low state, the virus is still effectively cleared, but cellular immunity $[C]$ and the inflammatory cytokine IL-6 rapidly decline to near-zero levels. This may reflect the action of immune suppression mechanisms or other immune modules (such as humoral and innate immunity) that effectively shut down immune responses after viral clearance, thereby preventing tissue damage caused by excessive immune activity. At the same time, innate immunity $[I]$ and IL-6 also decrease rapidly when cellular immunity $[C]$ is low, showing coordinated reduction. This indicates that strong cooperative regulatory mechanisms exist in the system, where the low activity of cellular immunity is further reinforced by negative feedback or immune suppression pathways, stabilizing the shutdown of immune responses. This state represents a more favorable recovery of immune homeostasis, suggesting that the host can return to normal levels quickly and reduce the risk of chronic inflammation or immune-related disorders.
In summary, the system exhibits a clear bistable behavior. This phenomenon reflects the diversity and complexity of interactions among immune cells and further demonstrates that the ultimate steady state of the immune system is strongly influenced by the initial immune conditions. These findings highlight the importance of early immune regulation in preventing long-term inflammation or chronic immune activation and provide theoretical support for the development of early intervention strategies in clinical practice.

To further analyze the sensitivity of immune system steady states to initial conditions, a state classification diagram was constructed with the initial level of innate immunity $I_0$ on the horizontal axis and the initial level of cellular immunity $C_0$ on the vertical axis, systematically examining the influence of different initial configurations on steady-state outcomes (Fig.~\ref{fig:result4}). Here, the viral clearance efficiency of cellular immunity $d_{V1}$ was fixed at a relatively high level to ensure that the virus could be cleared within a short time, thereby eliminating the interference of continuous viral input on immune steady states.

In the figure, red, blue, and green represent three typical combinations of steady-state responses. Red corresponds to the case where cellular immunity $[C]$ and the inflammatory cytokine IL-6 are both in high-expression states, indicating the possibility of immune overactivation or chronic inflammation. Blue denotes elevated IL-6 while cellular immunity $[C]$ remains in a low state, suggesting that inflammation is activated but cellular immunity is not effectively engaged. Green represents both variables at low-expression levels, corresponding to the ideal state in which the virus is completely cleared and the immune response is successfully shut down.

Figs.~\ref{fig:result4}(d)–(e) show the discrete distribution of final steady states under different combinations of initial values. The three types of steady states exhibit clearly defined boundaries in the initial state space (black dashed lines). However, in certain boundary regions, noticeable overlaps are observed, indicating that different steady states may arise from similar initial conditions. This phenomenon reflects the multistability characteristics of the system.

\begin{figure}[tbhp]
\begin{center}
\includegraphics[width=\textwidth]{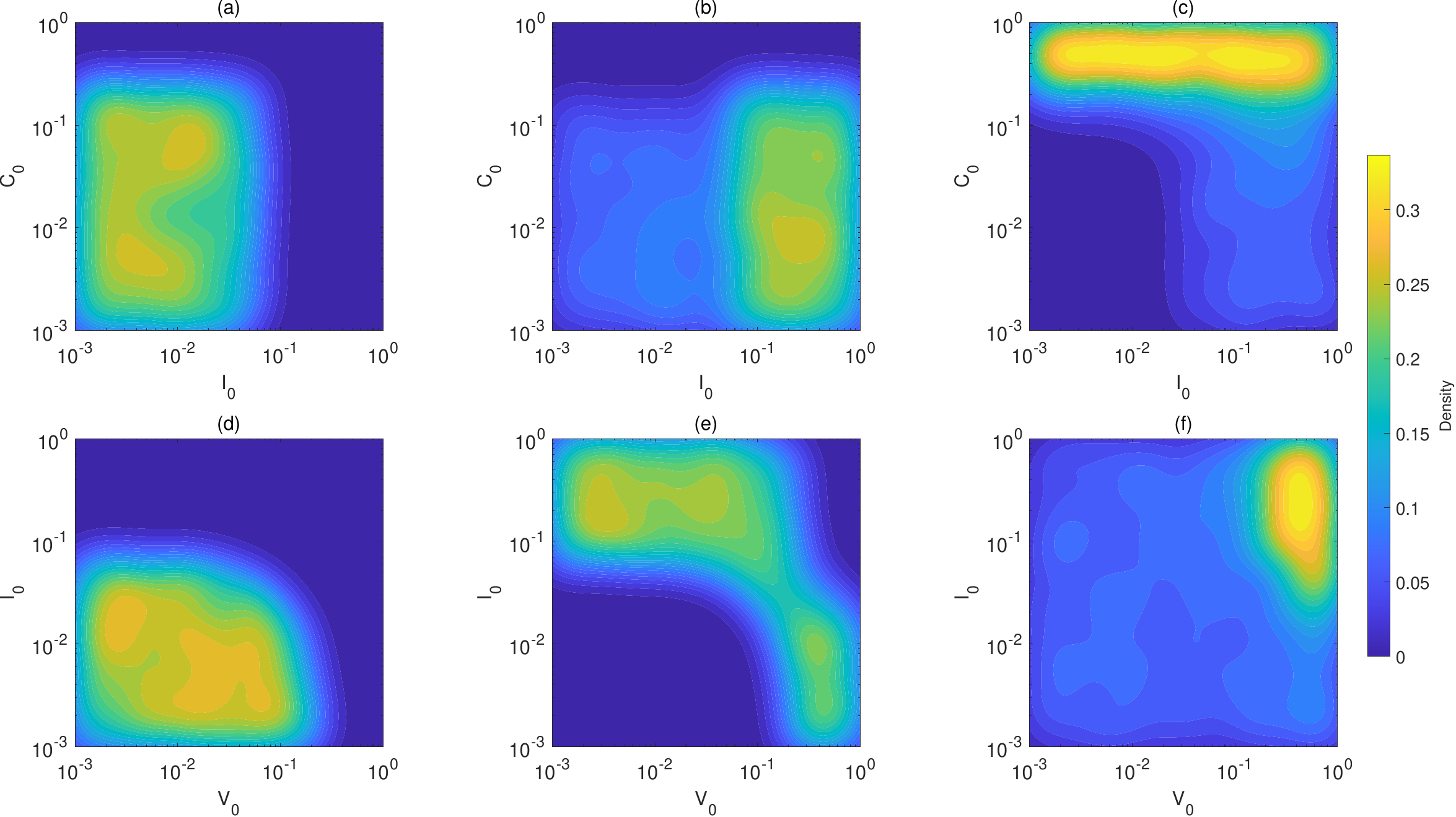}
\end{center}
\caption{Dependence of the density distribution of different steady states on initial conditions.
\textbf{(a)–(c)}  The density distribution maps based on Fig.~\ref{fig:result4}\textbf{(d)}, showing the distribution of different steady states under initial conditions $(I_0, C_0)$. \textbf{(d)–(f)} The density distribution maps based on Fig.~\ref{fig:result4}\textbf{(e)}, showing the distribution of steady states under initial conditions $(V_0, I_0)$. Each subplot illustrates the density with which the system converges to a specific immune steady state for different combinations of initial conditions, where brighter colors indicate higher probability of occurrence. 
These results highlight the multistability and sensitivity to initial conditions of the model.}\label{fig:tempt}
\end{figure}

To further characterize the distribution of different steady states in the initial condition space, we computed the occurrence frequency of each steady state on the parameter plane and presented the results as a density distribution map, as shown in Fig.~\ref{fig:tempt}. As a complement to the discrete classification results in Figs.~\ref{fig:result4}(d)–(e), this figure provides a more intuitive reflection of the probabilistic distribution of the three steady states in the boundary regions. The results show that while different steady states are clearly separated in most regions, significant overlaps and intersections occur near the critical boundaries. This indicates that the final outcome of the system is highly uncertain under such initial conditions, exhibiting typical features of multistability and sensitivity to initial conditions. In other words, the density distribution map provides a continuous probabilistic perspective for the discrete results, further highlighting the complex dynamical features in which the system may evolve into different steady states even under identical initial conditions.

This result reveals that when facing the same viral challenge, the long-term steady state of the immune system depends not only on system parameters but is also highly sensitive to initial conditions, showing typical nonlinear response and path-dependence characteristics. This implies that, in the early stages of disease progression, even minor differences in the immune system state—despite comparable viral loads—may lead to completely different inflammatory outcomes. Such findings are of great importance for identifying individual susceptibility, predicting immune outcomes, and developing early intervention strategies.

\subsection{Immune recovery dynamics and analysis of duration indicators under finite viral input}

\begin{equation}\label{eq:alpha}
\alpha(t) =
\begin{cases}
\phi, & t\in[0,T],\\[0.5ex]
0,    & t\in[T,\infty).
\end{cases}
\end{equation}

To investigate the recovery dynamics of the immune system under finite viral input, we simulated the case in which the virus was continuously introduced for $14 , \mathrm{units}$ and then stopped. As shown in Fig.~\ref{fig:result1}(b1), $\phi=0.01$, the viral load continuously increases and reaches a peak before the termination point of input at $t=14$. After the input ceases, the viral load rapidly declines and approaches zero, indicating that the virus is gradually cleared by the immune system.

In terms of the responses among different immune modules, innate immunity ($[I]$), cellular immunity ($[C]$), and humoral immunity ($[H]$) are all activated during the viral input period, showing a clear upward trend. Among them, humoral immunity ($[H]$) remains at a high level even after viral input ceases, then gradually declines and stabilizes, exhibiting a typical delayed-response characteristic.
As a representative marker of the inflammatory response, IL-6 rises synchronously during the viral peak but decreases much faster than the immune cell indicators. This phenomenon suggests that the inflammatory response decays rapidly during viral clearance, whereas the decline of immune cell activity shows a pronounced lag \cite{opsteen2023role}.

From the perspective of the overall system dynamics, the immune variables gradually return to steady states about $10 , \mathrm{units}$ after the termination of viral input ($t=24$), recovering to levels close to the initial immune state. This indicates that the immune system possesses strong self-regulatory capacity. However, the steady-state values of humoral immunity ($[H]$) and the immune suppression module ($[S]$) remain slightly higher than their initial levels and do not fully return.

This phenomenon can be explained from both the model mechanism and immunophysiological perspectives. From the viewpoint of dynamical structure, humoral immunity $[H]$ is jointly activated by cellular immunity ($[C]$) and innate immunity ($[I]$). After viral clearance, the residual activities of cellular immunity $[C]$ and innate immunity $[I]$ can still drive humoral immunity  humoral immunity $[H]$ to remain at a relatively high level for a short period. Together with the relatively small decay rate of humoral immunity $[H]$, this results in a delayed decline. Similarly, the production of immune suppression $[S]$ is driven by a constant term and immune activation variables (cellular immunity$[C]$ and innate immunity$[I]$), while being suppressed by the inflammatory factor IL-6. After viral clearance, the level of IL-6 rapidly decreases, weakening its inhibitory effect and further enhancing the net growth trend of immune suppression$[S]$, which ultimately causes its steady state to remain slightly higher than the initial level.

From an immunophysiological perspective, the residual activation of humoral immunity $[H]$ can be regarded as a “memory” response to potential reinvasion by pathogens, reflecting the persistence of humoral immunity. In contrast, the sustained high expression of immune suppression $[S]$ may correspond to the regulatory system’s braking mechanism that suppresses excessive immune responses and maintains system stability. Thus, although the immune system tends to stabilize as a whole, its recovery process exhibits clear module asynchrony and functional delay.
These results indicate that although viral clearance may be achieved within a short period after the termination of input, the full recovery of the immune system shows a pronounced delay effect, and the recovery speeds of different modules are not uniform. This dynamical feature is highly consistent with clinical observations of “residual immune activation during the recovery period after infection” \cite{opsteen2023role}.

\begin{figure}[tbhp]
\begin{center}
\includegraphics[width=\textwidth]{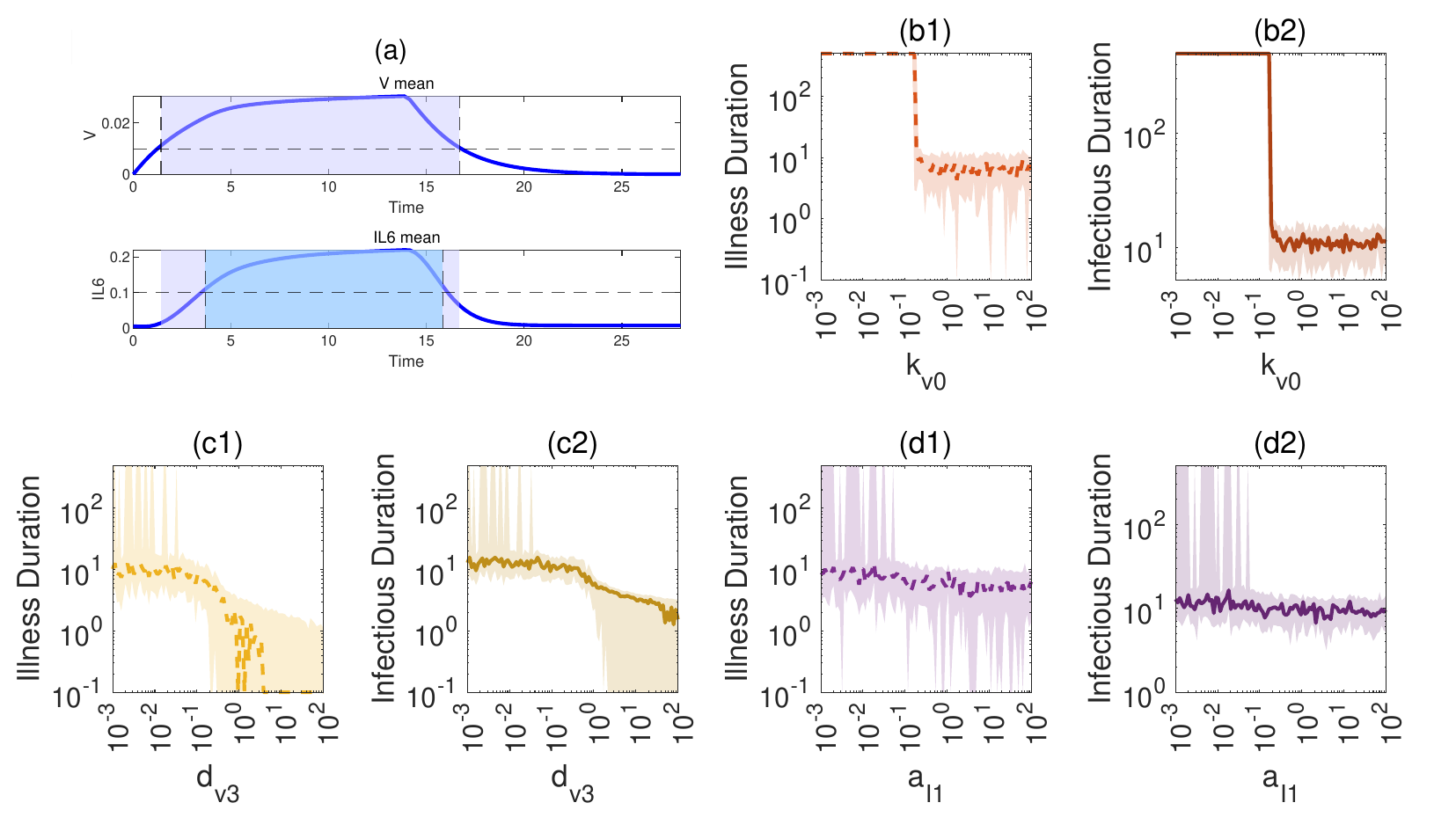}
\end{center}
\caption{
Analysis of the effects of parameter variations on the infection response dynamics of the immune system.\
\textbf{(a)} Definition of infectious duration and illness duration. The trajectories of viral load $[V](t)$ and pro-inflammatory cytokine $[IL-6](t)$ over time are shown. The blue solid line denotes the average responses of the two variables, while the shaded areas indicate the periods when infection is defined (virus load $[V(t)]$ or$[IL-6](t)$ above the thresholds $\theta_V$ and $\theta_{IL-6}$). Thresholds are set as $\theta_V=0.01$ and $\theta_{IL-6}=0.1$.
\textbf{(b1-b2)} Dependence of infectious duration and illness duration on the viral replication half-saturation constant $k_{v0}$. As $k_{v0}$ increases, the system exhibits a pronounced nonlinear transition around a critical value, rapidly shifting from a short-term infection state to a long-term infection state.
\textbf{(c1-c2)} Dependence of infectious duration and illness duration on the humoral immunity clearance parameter $d_{v3}$. When $d_{v3}$ is small, the disease duration is longer; as the parameter increases, the clearance efficiency of humoral immunity improves, leading to a gradual shortening of both disease and infectious durations, showing a continuous and smooth response.
\textbf{(d1-d2)} Dependence of infectious duration and illness duration on the innate immunity self-activation strength $a_{I1}$. When $a_{I1}$ is at a low level (around $10^{-2}$), disease duration exhibits instability and fluctuations. Infectious duration shows a similar unstable pattern, indicating that innate immunity self-activation plays a complex regulatory role in maintaining infection dynamics.
}\label{fig:result5}
\end{figure}

To quantify the continuous durations of different phases during infection, two temporal indicators were defined over the observation interval $[0, T_{\mathrm{end}}]$, based on the state trajectories of $[V]$ and IL-6 \cite{hadjichrysanthou2016understanding}. As shown in Fig.~\ref{fig:result5}(a), the thresholds $\theta_V$ and $\theta_{IL-6}$ represent the critical values for infectiousness and host inflammatory response, respectively.

\noindent\textbf{Infectious duration}
The onset time $t_{\mathrm{on}}^{(V)}$ is defined as the earliest time when $V(t)$ \emph{first} reaches or exceeds the threshold $\theta_V$. The offset time $t_{\mathrm{off}}^{(V)}$ is defined as the earliest time $t \geq t_{\mathrm{on}}^{(V)}$ when $V(t)$ \emph{first} falls below the threshold ($V(t)<\theta_V$). The infectious duration $D_{\mathrm{infec}}$ is then given by Eq.~\ref{D_infec}.
\begin{equation}
    D_{\mathrm{infec}} = t_{\mathrm{off}}^{(V)} -t_{\mathrm{on}}^{(V)} \label{D_infec}
\end{equation}

\noindent\textbf{Illness duration}
The onset time $t_{\mathrm{on}}^{(IL-6)}$ is defined as the earliest time when $IL\text{-}6(t)$ \emph{first} reaches or exceeds the threshold $\theta_{IL-6}$. The offset time $t_{\mathrm{off}}^{(IL-6)}$ is defined as the earliest time $t \geq t_{\mathrm{on}}^{(IL-6)}$ when $IL\text{-}6(t)$ \emph{first} falls below the threshold ($IL\text{-}6(t)<\theta_{IL-6}$). The illness duration $D_{\mathrm{ill}}$ is then given by Eq.~\ref{D_ill}.
\begin{equation}
    D_{\mathrm{ill}} = t_{\mathrm{off}}^{(IL6)} -t_{\mathrm{on}}^{(IL6)} \label{D_ill}
\end{equation}

The numerical solutions were obtained using the built-in MATLAB solver \texttt{ode15s}, and the onset/offset times were determined from the original non-uniform time points by detecting the \emph{first} crossing above or below the thresholds. To avoid spurious misjudgments caused by numerical noise, mild smoothing or a “minimum duration window” can be applied, without altering the main form of the above definitions.

To further reveal how illness duration and infectious duration respond to key system parameters, Fig.~\ref{fig:result5} illustrates the variations of these two dynamical indicators under multi-parameter perturbations. Fig.~\ref{fig:result5}(a) provides a schematic diagram of the infectious duration (gray shaded area) and the illness duration (blue shaded area), clearly defining the time intervals during which virus load $[V]$ and IL-6 exceed their respective thresholds, and thereby allowing the calculation of infectious duration and illness duration.
Fig.~\ref{fig:result5} also presents the response characteristics of key system variables under parameter perturbations, as well as their sensitivity to the two physiological indicators: “illness duration” and “infectious duration.” The overall figure consists of seven subplots, each depicting different aspects of the immune system’s dynamical response under parameter perturbations. Fig.~\ref{fig:result5}(a) shows a typical virus–inflammation dynamical process. The upper panel illustrates the temporal evolution of the average viral concentration $[V]$, while the lower panel shows the average level of the inflammatory factor IL-6 over time. The gray shaded region denotes the short-term infection phase, whereas the blue region denotes the phase of persistent infection or inflammatory activation. Around $t \approx 14 , \mathrm{units}$, the system undergoes viral input and reaches its peak, followed by transitions into different clearance or maintenance states. This figure is used to define the classification logic of the “illness period” and the “infectious phase”: virus load $[V]$ peaks and subsequently falls below a certain threshold, while IL-6 remains at a relatively high level for a sustained period.

Figs.~\ref{fig:result5}(b1)–(d2) present the variation curves of the key dynamical indicators—infectious duration and illness duration—with respect to different immune response–related parameters. In the calculations, each subplot examines the system’s response to the variation of a single parameter across different orders of magnitude, while all other parameters fluctuate randomly within $\pm 30\%$ probability. This design highlights both the independent effect of a specific parameter on system outputs and the intrinsic stochasticity of the system.

Specifically, to examine the effect of the viral self-replication half-saturation constant $k_{v0}$ on infectious and illness durations, parameter scanning was performed on $k_{v0}$. The results show that when $k_{v0}$ is small (approximately less than $0.1$), both infectious and illness durations are significantly shorter, indicating that viral replication can be effectively controlled by the immune system. However, once $k_{v0}$ exceeds a critical threshold, the infectious and illness durations increase rapidly and tend toward a stable chronic infection state. This dynamical response highlights the pronounced nonlinear sensitivity of the system to viral replication parameters, where even small parameter changes can trigger dramatic shifts in system dynamics (Figs.~\ref{fig:result5}(b1) and (b2)).

In addition, we examined the effect of the parameter $d_{v3}$, which represents the efficiency of virus clearance mediated by humoral immunity. As $d_{v3}$ increases, the ability of humoral immunity to clear the virus is enhanced, and both infectious duration and illness duration exhibit a gradually shortening trend. Unlike the sharp transitions induced by parameter $k_{v0}$, variations in $d_{v3}$ lead to smoother and more continuous system responses. This result indicates that although humoral immunity–mediated viral clearance has a noticeable impact on system behavior, it is not a key parameter driving system switches or abrupt transitions (Figs.~\ref{fig:result5}(c1) and (c2)).

Furthermore, to analyze the effect of the parameter $a_{I1}$, which represents the strength of innate immunity self-activation, we compared its variation with the two parameters discussed above. Overall, changes in $a_{I1}$ exert a milder influence on infectious duration and illness duration. However, when $a_{I1}$ is at relatively low levels (around $10^{-2}$), the system response still exhibits pronounced fluctuations and instability. This suggests that the self-feedback mechanism of innate immunity plays a strong nonlinear regulatory role in certain parameter regions, potentially inducing unstable infection states (Figs.~\ref{fig:result5}(d1) and (d2)).

In summary, the above analysis shows that parameters associated with different immune response mechanisms exert markedly different impacts on system dynamics. Among them, viral replication parameters (e.g., $k_{v0}$) play a dominant role in controlling the switches of infection states, whereas variations in the humoral immunity parameter ($d_{v3}$) and the innate immunity parameter ($a_{I1}$) respectively reflect the secondary regulatory roles of viral clearance and immune modulation mechanisms in controlling infection duration. Such complex multi-module interactions of the immune system indicate that its overall function depends not only on the coordination among modules but also on the timing and intensity of responses, thereby necessitating the consideration of dynamical sensitivity windows.

\subsection{Different regulatory relationships give rise to distinct forms of system dynamics}

Based on the above results, it can be seen that in addition to bistability, the system also exhibits pronounced oscillations and multistability under specific conditions \cite{novak2008design, ferrell2021systems}. Therefore, this section further focuses on the dynamical mechanisms underlying such complex behaviors. Specifically, we identify and analyze the core dynamical modules formed by the interactions among variables, and systematically investigate how key variable combinations influence these complex dynamical features.
To this end, subsystems of Eq.~\ref{model_equation} are studied separately, examining the interactions among variables within the system to gain deeper insight into the mechanisms by which oscillations arise.

First, considering that this study involves six key variables—virus ($[V]$), innate immunity ($[I]$), cellular immunity ($[C]$), humoral immunity ($[H]$), immune suppression ($[S]$), and the inflammatory factor (IL-6)—we introduce subsystems for analysis. Specifically, each subsystem consists of a subset of these six variables, while the remaining variables not included are fixed at their steady-state levels. In this way, different combinations of the six variables yield a total of $2^6 - 1 = 63$ possible subsystem structures.

For each subsystem structure, numerical simulations were performed using single-parameter variation. Specifically, in all subsystems one key parameter was varied while all other parameters were kept fixed, and the dynamic behaviors were examined for each subsystem structure in turn. This approach enables a systematic investigation of whether different subsystem combinations can give rise to bistability or other complex dynamic behaviors.

\begin{figure}[tbhp]
\begin{center}
\includegraphics[width=\textwidth]{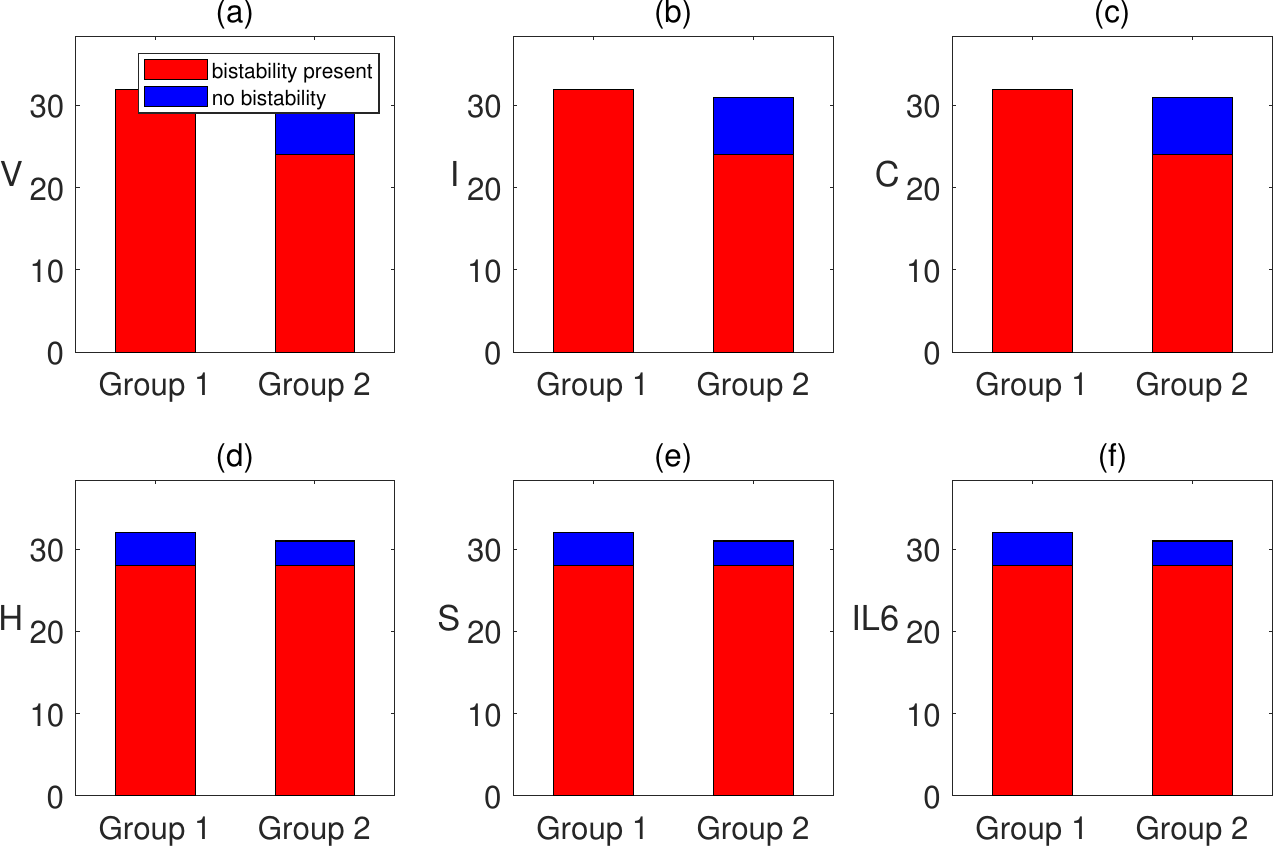}
\end{center}
\caption{Statistical analysis of the frequency of bistability occurrence in subsystems involving different variables. \textbf{(a)–(f)} Statistical analysis of different key variables ($[V](t)$, $[I](t)$, $[C](t)$, $[H](t)$, $[S](t)$, and $[IL-6](t)$), where the red regions indicate cases in which bistability is present, and the blue regions indicate cases in which bistability is absent. Group1 represents all subsystem configurations that include the given variable (a total of 32 cases), while Group2 represents all subsystem configurations that exclude the given variable (a total of 31 cases, excluding the trivial case in which no variable is present). This statistical analysis reveals the relative importance and sensitivity of different variables in contributing to bistable behavior of the system.
}\label{fig:fig_hist}
\end{figure}

As shown in Fig.~\ref{fig:fig_hist}, we performed a frequency analysis of bistability occurrence in subsystems involving each variable. Group1 represents all subsystem configurations that include the given variable, while Group~2 represents those that exclude it. Red indicates the frequency of bistability occurrence in subsystems involving the variable, whereas blue indicates the absence of bistability. The analysis reveals that different variables contribute to the emergence of complex system behaviors to varying degrees of importance. For example, subsystems involving virus ($[V]$), innate immunity ($[I]$), and cellular immunity ($[C]$) exhibit a notably higher proportion of bistability, suggesting that the interactions among these three variables are critical for the emergence of complex dynamic behaviors. In contrast, subsystems involving immune suppression ($[S]$), humoral immunity ($[H]$), and the inflammatory factor (IL-6) show relatively lower frequencies of bistability, indicating that interactions among these variables alone are insufficient to robustly generate complex dynamic features.

In summary, the above analysis identifies virus ($[V]$), innate immunity ($[I]$), and cellular immunity ($[C]$) as the core variable combination, which plays a critical role in shaping the system’s complex dynamic behaviors. This finding lays a solid foundation for further in-depth investigation of the dynamic mechanisms underlying virus–immune responses and provides a theoretical basis for the design of clinical intervention strategies.

\section{Conclusion}
  
This study is based on a six-variable ordinary differential equation model incorporating viral load ($[V]$), innate immunity ($[I]$), cellular immunity ($[C]$), humoral immunity ($[H]$), immune suppression ($[S]$), and the inflammatory cytokine IL-6, to systematically investigate the nonlinear dynamic features of virus–immune system interactions. The model employs Hill functions to uniformly capture nonlinear activation, inhibition, and saturation effects across modules \cite{alon2019introduction}, and explicitly introduces an external input $\alpha(t)$ to characterize both continuous and short-term viral exposure scenarios. In this way, the immune network’s global response can be quantitatively assessed within a unified framework.

Bifurcation analysis of the model shows that, under continuous input, the relative magnitudes of viral replication rate and immune clearance efficiency directly determine the system’s steady-state landscape. When replication is strong and clearance is insufficient, the system may remain in a stable high-virus–high-inflammation state; when clearance is enhanced, the system can switch to a healthy low-virus–low-inflammation state. Two-dimensional parameter scans further reveal that, under different parameter combinations, the system may exhibit monostability, coexistence of multiple steady states, or even strong sensitivity to initial conditions. This implies that even with similar exposure and parameter settings, long-term immune outcomes may differ substantially across individuals \cite{eftimie2016mathematical}. Moreover, we identified the coexistence of “high/low states”: even when viral levels are close to zero, cellular immunity and IL-6 can stabilize at distinct levels, indicating that positive and negative feedback can sustain multiple stable states under low viral pressure.
Under short-term input conditions, the system exhibits typicalmulti-timescale immune recovery dynamics, viral load declines rapidly once input ceases, while IL-6 simultaneously decays quickly, reflecting the transient nature of the inflammatory response. In contrast, humoral immunity and the immune suppression module remain elevated for a longer period and decrease more slowly, showing residual effects and asynchronous recovery. To quantitatively characterize this process, we defined the “infectious duration” and “illness duration” based on viral load and IL-6 thresholds, and examined their sensitivity under multiparameter perturbations. The results indicate that viral replication–related parameters play a dominant role in controlling the course of infection, capable of triggering nonlinear transitions from acute to chronic states, whereas humoral clearance efficiency and innate self-activation strength act as modulators, influencing the length and stability of infection and inflammation processes. These findings reveal that immune recovery after viral clearance is inherently unbalanced and delayed, providing a theoretical explanation for the clinically observed phenomenon of “residual immune activation during recovery.”
Subsystem enumeration analysis further shows that the $[V]$–$[I]$–$[C]$ triplet is the core structure driving multistability and complex oscillatory behavior, while $[H]$ and the $[S]$–IL-6 module primarily affect the amplitude of responses and the synchronicity of recovery. This finding provides a mechanistic basis for model reduction and the identification of potential intervention targets.

In summary, the mesoscale immune dynamics framework proposed in this study reveals how the competition between viral replication and immune clearance shapes multistability and sensitivity to initial conditions through bifurcation mechanisms, and highlights the temporal differences among distinct immune modules during recovery. In addition, we introduced temporal indicators to compare system outcomes under different parameter settings and input modes. The framework advances theoretical understanding of virus–immune system coupling, provides methodological tools for parameter analysis and threshold determination, and offers practical insights for immune intervention studies. Future work may incorporate individualized longitudinal data for parameter fitting, extend the framework to models with stochasticity and spatial structures, and explore regulatory strategies for key parameters, thereby promoting the integration of mathematical modeling with clinical immunology.

\bibliographystyle{plain}  


\end{CJK*}
\end{document}